\documentclass[12pt,epsf]{article}
\usepackage{graphicx,epsf}
\usepackage{lscape}
\usepackage{citesort}
\begin{document}

\def\ds{\displaystyle}
\def\beq{\begin{equation}}
\def\eeq{\end{equation}}
\def\bea{\begin{eqnarray}}
\def\eea{\end{eqnarray}}
\def\beeq{\begin{eqnarray}}
\def\eeeq{\end{eqnarray}}
\def\ve{\vert}
\def\vel{\left|}
\def\ver{\right|}
\def\nnb{\nonumber}
\def\ga{\left(}
\def\dr{\right)}
\def\aga{\left\{}
\def\adr{\right\}}
\def\lla{\left<}
\def\rra{\right>}
\def\rar{\rightarrow}
\def\nnb{\nonumber}
\def\la{\langle}
\def\ra{\rangle}
\def\ba{\begin{array}}
\def\ea{\end{array}}
\def\tr{\mbox{Tr}}
\def\ssp{{\Sigma^{*+}}}
\def\sso{{\Sigma^{*0}}}
\def\ssm{{\Sigma^{*-}}}
\def\xis0{{\Xi^{*0}}}
\def\xism{{\Xi^{*-}}}
\def\qs{\la \bar s s \ra}
\def\qu{\la \bar u u \ra}
\def\qd{\la \bar d d \ra}
\def\qq{\la \bar q q \ra}
\def\gGgG{\la g^2 G^2 \ra}
\def\q{\gamma_5 \not\!q}
\def\x{\gamma_5 \not\!x}
\def\g5{\gamma_5}
\def\sb{S_Q^{cf}}
\def\sd{S_d^{be}}
\def\su{S_u^{ad}}
\def\ss{S_s^{??}}
\def\sbp{{S}_Q^{'cf}}
\def\sdp{{S}_d^{'be}}
\def\sup{{S}_u^{'ad}}
\def\ssp{{S}_s^{'??}}
\def\sig{\sigma_{\mu \nu} \gamma_5 p^\mu q^\nu}
\def\fo{f_0(\frac{s_0}{M^2})}
\def\ffi{f_1(\frac{s_0}{M^2})}
\def\fii{f_2(\frac{s_0}{M^2})}
\def\O{{\cal O}}
\def\sl{{\Sigma^0 \Lambda}}
\def\es{\!\!\! &=& \!\!\!}
\def\ap{\!\!\! &\approx& \!\!\!}
\def\ar{&+& \!\!\!}
\def\ek{&-& \!\!\!}
\def\kek{\!\!\!&-& \!\!\!}
\def\cp{&\times& \!\!\!}
\def\se{\!\!\! &\simeq& \!\!\!}
\def\eqv{&\equiv& \!\!\!}
\def\kpm{&\pm& \!\!\!}
\def\kmp{&\mp& \!\!\!}

% .........................................................

\def\simlt{\stackrel{<}{{}_\sim}}
\def\simgt{\stackrel{>}{{}_\sim}}

% .........................................................

\title{ {\Large {\bf Semileptonic $D_{q}\rightarrow
K_{1}\ell \nu$ and nonleptonic $D\rightarrow K_1 \pi$ decays in
three--point QCD sum rules and factorization approach } } }

\author{\vspace{1cm}\\
{\small R. Khosravi$^1$ \thanks {e-mail: khosravi.reza @
gmail.com}~\,}, {\small K. Azizi $^2$\thanks {e-mail: e146342 @
metu.edu.tr}~\,}, {\small N. Ghahramany$^1$ \thanks
{e-mail:ghahramany @ susc.ac.ir
}~\,}\\
 {\small $^1$ Physics Department , Shiraz University, Shiraz 71454,
Iran}\\
{\small  $^2$ Department of Physics, Middle East Technical
University, 06531 Ankara, Turkey}\\}
\date{}

\begin{titlepage}
\maketitle \thispagestyle{empty}

\begin{abstract}
We analyze the semileptonic $D_{q}\to K_1 \ell\nu$ transition with
$q=u,~ d, ~s$,
 in the framework of the three--point QCD sum rules and
the nonleptonic $D\to K_1 \pi$ decay  within the QCD  factorization
approach. We study $D_{q}$ to $K_1(1270)$ and $K_1(1400)$ transition
form factors by separating the mixture of  the $K_1(1270)$ and
$K_1(1400)$ states. Using the transition form factors of the $D\to
K_1 $, we analyze the nonleptonic $D\to K_1 \pi$ decay. We also
present the decay amplitude and decay width of these decays in terms
of the transition form factors. The branching ratios of these channel modes
are also calculated at different values of the mixing angle
$\theta_{K_1}$ and compared with the existing experimental data for the nonleptonic case.
\end{abstract}
%\vspace{1cm}

\end{titlepage}

\section{Introduction}
Analyzing  the semileptonic decays of the charmed $D_{q}$ mesons is
very useful for determination of the elements of
the Cabibbo-Kabayashi-Maskawa (CKM) matrix and also leptonic decay
constants of the initial and final meson states. The semileptonic
 $D_{s}\to K_1 \ell\nu$ transition could give useful
information about the internal structure of the $D_{s}$ meson.
Investigating the nonleptonic decays such as $D\to K_1 \pi$ can also
be important for interpretation of the structure of the lightest
scaler mesons \cite{Bediaga}.

From the experimental view, the physical states $K_1(1270)$ and
$K_1(1400)$ are the mixtures of the strange members of two
axial-vector $SU(3)$ octets $1^3P_1(K_{1A})$ and $1^1P_1(K_{1B})$.
The $K_{1A}$ and $K_{1B}$ are not mass eigenstates and they can be
mixed together due to the nonstrange light quark mass difference.
Their relations with the $K_1(1270)$ and $K_1(1400)$ states can be
written as \cite{Lee,Suzuki,kazem1}:

\begin{eqnarray}\label{eq1}
\mid K_1(1270)>&=&\mid K_{1A}> sin\theta_{K_1}~+\mid
K_{1B}>cos\theta_{K_1}, \nonumber\\
\mid K_1(1400)>&=&\mid K_{1A}> cos\theta_{K_1}~-\mid
K_{1B}>sin\theta_{K_1}.
\end{eqnarray}

The angle $\theta_{K_1}$ has been obtained with two-fold ambiguity
$\mid \theta_{K_1} \mid \approx 33^\circ$, as given in Ref
\cite{Suzuki}. Also in Ref \cite{Burakovsky} $ 35^\circ \leq \mid
\theta_{K_1} \mid \leq 55^\circ$ has been found. In this paper we
use $\theta_{K_1}$ in the interval $ 37^\circ\leq \mid
\theta_{K_1} \mid \leq 58^\circ$ \cite{kazem1,Cheng2}. The sign ambiguity
for $\theta_{K_1}$ is due to the fact that one can add arbitrary
phases to $\mid K_{1A}>$ and $\mid K_{1B}>$ states.

The QCD sum rules approach has been successfully applied to a wide variety
of problems in charm meson decays. The semileptonic decays $D_s\to f_0
\ell \nu$, $D_s\to \phi \ell \nu$ \cite{Bediaga}, $D\to
\overline{K}^{0} \ell \nu$ \cite{Aliev}, $D^+\to K^{0*} e^+
\nu_{e}$ \cite{Ball1}, $D\to \pi \ell \nu$ \cite{Ball2}, $D\to
\rho \ell \nu$ \cite{Ovchinnikov}, $D^+_s\to \phi \bar{\ell} \nu$
\cite{Du} and $D\to K_0^* \bar{\ell} \nu$ \cite{Yang} have been studied
in the framework of the three--point QCD sum rules. As a  nonperturbative method, the QCD sum rules has been of interest and it is a well established technique in the hadron physics since it is based on the fundamental QCD Lagrangian (for details about the QCD sum rules approach see for instance \cite{colangelo}).

In the present work, we study the semileptonic decays of the
$D_{q}\to K_1 \ell \nu$  in the framework of the three--point QCD
sum rules. The long distance dynamics of such transitions can be
parameterized in terms of some form factors calculating of which
play fundamental role in the analyzing of such type transitions.
Considering   the contributions of the operators with mass dimension
$d=3, 4, 5$ as condensate and non-perturbative contributions, first
we calculate the transition form factors of the semileptonic $D_q\to
K_1 \ell \nu (q=u, d, s)$ decays. Using these form factors, the
total decay width as well as the branching ratio for  the
aforementioned transitions are also evaluated at different values of
the mixing angle. Having computed the form factors of the $D\to
K_1$, the amplitude and decay rate of the  nonleptonic $D_{u,d}\to
K_1 \pi$ decays are also computed in terms of  those form factors
using the QCD  factorization method (for more  about the method see
\cite{beneke1,beneke2,beneke3} and references therein).

The paper is organized as follows. The calculation of the sum rules
for the relevant form factors are presented in section2. In
calculating the form factors, first we consider the general $\langle
K_{1}|$ state. Then, using the definition of the G-parity conserving
decay constant $<0\mid J^{\nu}_{K_{1A}} \mid
K_{1A}(p',\varepsilon)>=f_{K_{1A}}m_{K_{1A}}\varepsilon^{\nu}$ and
G-parity violating decay constant $ <0\mid J^{\nu}_{K_{1B}} \mid
K_{1B}(p',\varepsilon)>=f_{K_{1B^{\perp}}}(1~~GeV)a_{0}^{\parallel,K_{1B}}
m_{K_{1}^{B}}\varepsilon^{\nu}$, where $a_{0}^{\parallel,K_{1B}}$ is
the zeroth Gegenbauer moment of $K_{1B}$ state and it is zero in the
$SU(3)$ symmetry limit, we obtain  the form factors of the $D\to
K_{1A(B)}$ states. Finally, considering  Eq. (\ref{eq1}), we
separate the $\langle K_{1}[1270(1400)]|$ states and derive   form
factors of the $D\to K_{1}[1270(1400)]$ transitions.   The decay
rate formulas for semileptonic and nonleptonic cases are presented
in section3. We derive the decay rate formula for $D\to K_1\pi$
decay using the QCD factorization method in tree level. Section 4 is
devoted to the numeric analysis of the form factors as well as the
branching fractions of the considered semileptonic and non-leptonic
decays at different values of the mixing angle,   and discussions. A
comparison of our results for  the branching ratios for the
non-leptonic case  with the existing  experimental data is also made
in this section.

\section{Sum rules for $D_{q}\rightarrow K_{1}\ell \nu$ transition form factors}

The $D_{q}\rightarrow K_{1} \ell \nu$ with $q=u, d, s$ decay
governed by the tree level $c\rightarrow q^{\prime}$ ($q^{\prime}=d, s)$
transition  (see Fig .1).
\begin{figure}[th]

\vspace*{4.cm}
\begin{center}
\begin{picture}(160,10)
\centerline{ \epsfxsize=16cm \epsfbox{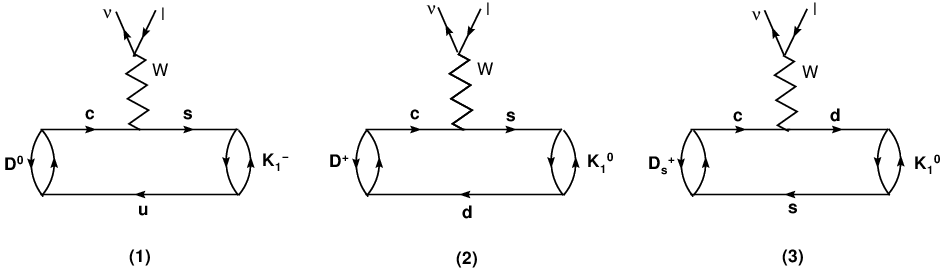}}
\end{picture}
\end{center}

\vspace*{0cm} \caption{Semileptonic decays of $D_{q}$ to $K_1$.
Diagrams 1, 2 and 3 are related to the $D^0\to K_1^- \ell \nu$,
$D^+\to K_1^0 \ell \nu$ and  $D^+_s\to K_1^0 \ell \nu$,
respectively.}
\end{figure}
\normalsize

\setlength{\unitlength}{1mm}

In the standard model, the effective Hamiltonian responsible for
these transitions is given as:
\begin{equation}  \label{eq2}
\mathcal{H}_{eff}=\frac{G_{F}}{\sqrt{2}} V_{cq^{'}}~\overline{\nu}
~\gamma_{\mu}(1-\gamma_{5})~l~\overline{q}^{'} ~\gamma_{\mu}(1-\gamma_{5}) c, \\
\end{equation}
where, $G_{F}$ is the Fermi constant and $V_{cq^{'}}$ are the CKM
matrix elements. The decay amplitude for $D_{q}\rightarrow K_{1}
\ell \nu$ is obtained by inserting  Eq. (\ref{eq2}) between the
initial and final meson states.
\begin{equation}  \label{eq3}
\mathcal{M}=\frac{G_{F}}{\sqrt{2}} V_{cq^{'}}~\overline{\nu} ~\gamma_{\mu}(1-%
\gamma_{5})l< K_{1}(p^{\prime},\varepsilon)\mid~\overline{q}^{'} ~\gamma_{\mu}(1-%
\gamma_{5}) c\mid D_{q}(p)>.
\end{equation}

The next step is to calculate the matrix element appearing in Eq.
(\ref{eq3}). Both axial and vector parts of the transition current
give contribution to this matrix element and it can be parametrized
in terms of some form factors  using the  Lorentz invariance and
parity conservation as follows:
\begin{equation}  \label{eq4}
<K_{1}(p^{\prime},\varepsilon)\mid\overline{q}^{'}\gamma_{\mu}\gamma_{5}
c\mid D_{q}(p)>=-\frac{2f_V^{D_{q}\to K_1}(q^2)}{(m_{D_q}+m_{K_1})}\epsilon_{\mu\nu\alpha\beta} \varepsilon^{\nu}p^\alpha
{p^{\prime}}^\beta,
\end{equation}
\begin{eqnarray}  \label{eq5}
<K_{1}(p^{\prime},\varepsilon)\mid\overline{q}^{'}\gamma_{\mu}
c\mid D_{q}(p)> &=&i\left[f_{0}^{D_{q}\to
K_1}(q^2)(m_{D_q}+m_{K_1})\varepsilon_{\mu} \right.
\nonumber \\
-\frac{f_{1}^{D_{q}\to
K_1}(q^2)}{(m_{D_q}+m_{K_1})}(\varepsilon p)P_{\mu} &-&
\left. \frac{f_{2}^{D_{q}\to K_1}(q^2)}{(m_{D_q}+m_{K_1})}(\varepsilon p)q_{\mu}%
\right].
\end{eqnarray}
In order for the calculations to be simple,  the following
redefinitions are used
\begin{eqnarray}  \label{eq6}
F_{V}^{D_{(s)}\to K_1}(q^2)&=&\frac{2f_V^{D_{q}\to K_1}(q^2)}{(m_{D_q}+m_{K_1})}%
~,~~~~~~~~~~~~F_{0}^{D_{q}\to K_1}(q^2)=f_{0}^{D_{q}\to
K_1}(q^2)(m_{D_q}+m_{K_1}), \nonumber
\\
F_{1}^{D_{q}\to K_1}(q^2)&=&-\frac{f_{1}^{D_{q}\to
K_1}(q^2)}{(m_{D_q}+m_{K_1})}~,~~~~~~~~~~~~ F_{2}^{D_{q}\to
K_1}(q^2)=-\frac{f_{2}^{D_{q}\to K_1}(q^2)}{(m_{D_q}+m_{K_1})},
\end{eqnarray}
where the $F_V^{D_{q}\to K_1}(q^2)$, $F_{0}^{D_{q}\to
K_1}(q^2)$, $F_{1}^{D_{q}\to K_1}(q^2)$ and $F_{2}^{D_{q}\to
K_1}(q^2)$ are
the new transition form factors,  $P_{\mu}=(p+p^{\prime})_{\mu}$, $%
q_{\mu}=(p-p^{\prime})_{\mu}$ and $\varepsilon$ is the
four--polarization vector of the axial  $K_1$ meson.

Based on the general philosophy of the three-point QCD sum rules technique, the above form factors in
Eq. (\ref{eq6}) can be evaluated from the time ordered product of
the following three currents.
\begin{eqnarray}\label{eq7}
\Pi^{(V-A)}_{\mu\nu} (p^2,p'^2,q^2) = i^2 \int d^4 x d^4 y e^{+i p^\prime
x - i p y} \left< 0 \left| \mbox{\rm T} \left\{ J_{K_{1}\nu}(x)
J^{(V-A)}_\mu (0)  J^{\dagger}_{D_q}(y)\right\} \right| 0\right>~,
\end{eqnarray}
where,
$J_{K_{1}\nu}(x)=\overline{q_{1}}\gamma_{\nu}\gamma_{5}s$ ($q_{1}=u,d$) ,
 $J_{D_{q}}(y)=\overline{q}\gamma_{5}c$ are the interpolating currents of the $K_1^{0-}$ and $D_{q}$ and  $J_{\mu}^{V}=\overline{q^{'}}\gamma_{\mu}c$ and
 $J_{\mu}^{A}=\overline{q^{'}}\gamma_{\mu}\gamma_{5}c$ are the  vector and axial-vector parts of the transition current, respectively.

The above correlation function is calculated in two different
approaches: On the quark level, it describes a meson as quarks and
gluons interacting in a QCD vacuum. This is called the theoretical
or QCD side. In the phenomenological or physical side, it is
saturated by a tower of mesons with the same  quantum numbers as the
interpolating currents. The  form factors are determined by matching
these two different representations of the correlation function and
applying double Borel transformation with respect to the momentum of
the initial and final meson states to suppress the contribution
coming from the higher states and continuum. We can express the
correlation function in both sides in terms of four independent
Lorentz structures:
\begin{eqnarray}\label{eq8}
\Pi_{\mu\nu}^{(V-A)} = \epsilon_{\mu\nu\alpha\beta} \,p^\alpha
p^{\prime\beta} \Pi_{V}+ g_{\mu\nu} \Pi_{0}+ P_\mu p_\nu \Pi_{1}+
q_\mu p_\nu \Pi_{2}.
\end{eqnarray}
To find the sum rules for the related form factors, we will match the coefficients of the corresponding structures from both representations of the correlation function.

First, we calculate the aforementioned correlation function in the phenomenological representation. Inserting two complete sets of intermediate states with the
same quantum number as the currents $J_{K_1}$ and $J_{D_q}$ to Eq. (\ref{eq7}), we
obtain
\begin{eqnarray} \label{eq9}
&&\Pi _{\mu\nu}^{V-A}(p^2,p'^2,q^2)=
\nonumber \\
&& \frac{<0\mid J_{K_1 \nu} \mid K_1(p',\varepsilon)><
K_1(p',\varepsilon)\mid J_{\mu}^{V-A}\mid D_q(p)><D_q(p)\mid
J^{\dag}_{D_q}\mid0>}{(p'^2-m_{K_1}^2)(p^2-m_{D_q}^2)}+
\nonumber\\
&&\mbox{ the  higher resonances and continuum.}
\end{eqnarray}

In Eq. (\ref{eq9}), the vacuum to initial and final meson states
matrix elements are defined  as:
\begin{eqnarray}  \label{eq10}
<0\mid J_{K_{1}}^{\nu}\mid K_1(p^{\prime
})>=f_{K_1}m_{K_1}\varepsilon ^{\nu }~,~~
<0\mid J_{D_q}\mid D_{q}(p)>=i\frac{f_{D_q%
}m_{D_q}^{2}}{m_{c}+m_{q}},
\end{eqnarray}
where $f_{K_1}$ and $f_{D_q}$ are the leptonic decay constants of
$K_1$ and $D_q$ mesons, respectively. Using Eq. (\ref{eq4}), Eq.
(\ref{eq5}) and Eq. (\ref{eq10}) in Eq. (\ref{eq9}) and performing
summation over the polarization vector of the  $K_1$ meson, we get the
following result for the physical part:
\begin{eqnarray}\label{eq11}
\Pi_{\mu\nu}^{V-A}(p^2,p'^2,q^2)&=&-\frac{f_{D_q}m_{D_q}^2}{(m_{c}+m_{q})}\frac{f_{K_1}m_{K_1}}
{(p'^2-m_{K_1}^2)(p^2-m_{D_q}^2)} \times [F_{0}^{D_{(s)}\to
K_1}(q^2)g_{\mu\nu}\nonumber \\&+&F_{1}^{D_{(s)}\to
K_1}(q^2)P_{\mu}p_{\nu}+F_{2}^{D_{(s)}\to
K_1}(q^2)q_{\mu}p_{\nu}+i~F_V^{D_{(s)}\to K_1}(q^2)
\epsilon_{\mu\nu\alpha\beta}p'^{\alpha}p^{\beta}]\nonumber
\\&+& \mbox{excited states.}
\end{eqnarray}
The coefficients of the Lorentz structures
$i\epsilon_{\mu\nu\alpha\beta}p^{\alpha}p^{'\beta}$, $g_{\mu\nu}$,
$P_{\mu}p_{\nu}$ and $q_{\mu}p_{\nu}$ in the correlation function
$\Pi_{\mu}^{V-A}$ will be chosen  in determination of the form
factors $F_V^{D_{(s)}\to K_1}(q^2)$,~$F_{0}^{D_{(s)}\to
K_1}(q^2)$,\\$F_{1}^{D_{(s)}\to K_1}(q^2)$ and $F_{2}^{D_{(s)}\to
K_1}(q^2)$, respectively.

On the QCD or theoretical side, the correlation function is calculated in the quark and gluon languages by the help of  the operator product expansion (OPE) in the deep Euclidean
region where $p^2\ll (m_c+m_q)^2$, ${p^{'}}^2\ll
(m_q^2+m_{q^{'}}^2)$. In Eq. (\ref{eq7}), using the expansion of the time
ordered products of the currents,  the three--point
correlation function is written in terms of the series of local operators with
increasing dimension as the following form
 \cite{Aliev2}:
\begin{eqnarray}\label{eq12}
-\int d^4x d^4y e^{i(p x - p^\prime y)} T \Big\{J_{K_1\nu} J_\mu
J^{\dag}_{D_q} \Big\} \es (C_0)_{\mu\nu} I + (C_3)_{\mu\nu}
\overline{\Psi} \Psi + (C_4)_{\mu\nu} G_{\alpha\beta}
G^{\alpha\beta}\ra \nnb \\
\ar (C_5)_{\mu\nu} \overline{\Psi} \sigma_{\alpha\beta}
G^{\alpha\beta} \Psi + (C_6)_{\mu\nu} \overline{\Psi} \Gamma \Psi
\overline{\Psi} \Gamma^\prime \Psi~,
\end{eqnarray}
where, $G_{\alpha\beta}$ is the gluon field strength tensor, $(C_i)_{\mu\nu}$ are the Wilson coefficients,
 $I$ is the
unit matrix, $\Psi$ is the local  field operator of the light
quarks, and $\Gamma $ and $\Gamma^{'}$ are the matrices appearing in
the calculations. Taking into account the vacuum expectation value
of the OPE, the expansion of the correlation function in terms of
the local operators is written as follows:
\begin{eqnarray}\label{eq13}
\Pi_{\mu\nu} (p_1^2,p_2^2,q^2) \es C_{0\mu\nu} + C_{3\mu\nu} \la
\overline{\Psi} \Psi \ra + C_{4\mu\nu} \la G^2 \ra + C_{5\mu\nu}
\la \overline{\Psi}
\sigma_{\alpha\beta} G^{\alpha\beta} \Psi \ra \nnb \\
\ar C_{6\mu\nu} \la \overline{\Psi} \Gamma \Psi \overline{\Psi}
\Gamma^\prime \Psi \ra~.
\end{eqnarray}

In  Eq.(\ref{eq13}),  the contributions of the perturbative and
condensate terms of dimension $3, 4$, and $ 5$ as non-perturbative
parts are considered. The diagrams for the contributions of the
non-perturbative part are depicted in Figs. 2, 3 and 4. It's found
that the heavy quark condensate contributions are suppressed by
inverse of the heavy quark mass and can be safely removed (see
diagrams 4, 5, 6 in Fig. 2). The light $q^{'}$ quark condensate
contributions are zero after applying the double Borel
transformation with respect to  both variables $p^2$ and ${p^{'}}^2$
since only one variable appears in the denominator (see diagrams 1,
2, 3 in Fig. 2).
\begin{figure}[th]

\vspace*{2.cm}
\begin{center}
\begin{picture}(160,80)
\centerline{ \epsfxsize=12cm \epsfbox{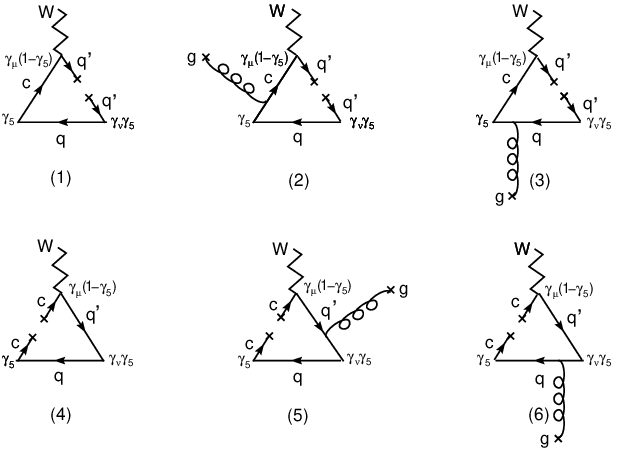}}
\end{picture}
\end{center}

\vspace*{0cm} \caption{The  quark condensate
diagrams  without any gluon and with   one gluon emission.}
\normalsize
\end{figure}
\setlength{\unitlength}{1mm}

Our calculations show that in this case, the two-gluon condensate
contributions (see diagrams in Fig. 3) are very small in comparison
with the quark condensate contributions and we can easily ignore
their contributions in our calculations.
\\
\\
\\

\begin{figure}[th]

\vspace*{2.cm}
\begin{center}
\begin{picture}(160,50)
\centerline{ \epsfxsize=12cm \epsfbox{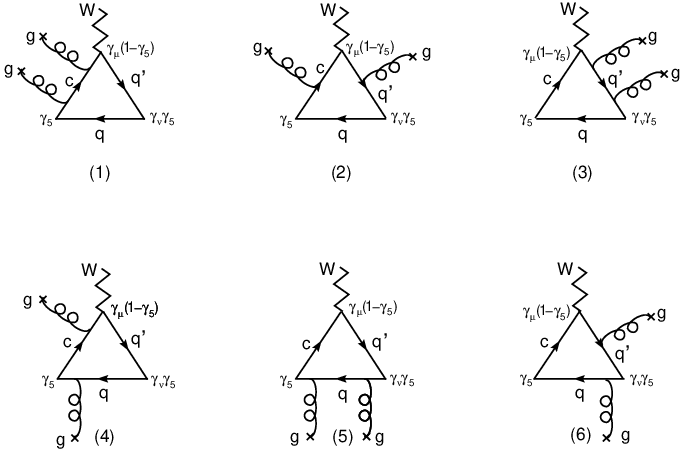}}
\end{picture}
\end{center}

\vspace*{0cm} \caption{Diagrams for two-gluon condensate
contributions.}
\end{figure}
\normalsize

\setlength{\unitlength}{1mm}

Therefore, the main contribution in the non-perturbative part comes
from the q-quark condensates. (see Fig. 4).
\begin{figure}[th]

\vspace*{3.cm}
\begin{center}
\begin{picture}(160,10)
\centerline{ \epsfxsize=12cm \epsfbox{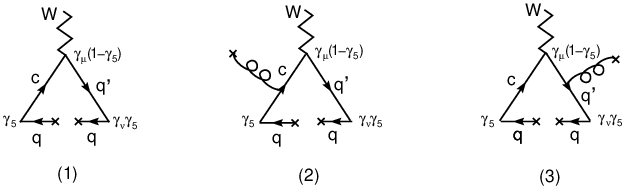}}
\end{picture}
\end{center}

\vspace*{0cm} \caption{Diagrams for q-quark condensates
contributions. }
\end{figure}
\normalsize

\setlength{\unitlength}{1mm} As a result, in the lowest order of the
perturbation theory, the three--point correlation function receives
a contribution from the perturbative part ( bare-loop contributions
of diagrams in Fig. 1) and nonp-erturbative part (contributions of
the diagrams shown in  Fig. 4) i.e.,
\begin{eqnarray}\label{eq14}
\Pi_i(p^2,{p^{'}}^2,q^2) = \Pi_i^{per}(p^2,{p^{'}}^2,q^2) +
\Pi_i^{non-per} (p^2,{p^{'}}^2,q^2) ~.
\end{eqnarray}

Using the double dispersion representation, the bare-loop
contribution is determined:
\begin{eqnarray}\label{eq15}
\Pi_i^{per} = - \frac{1}{(2 \pi)^2} \int \int\frac{\rho_i^{per}
(s,s^\prime, q^2)}{(s-p^2) (s^\prime - p^{\prime 2})} ds ds^\prime +
\mbox{\rm subtraction terms}~,
\end{eqnarray}

The following inequality  is responsible  for obtaining  the
integration limits in  Eq. (\ref{eq15}).
\begin{eqnarray}\label{eq16}
-1 \le \frac{2 s s^\prime + (s + s^\prime - q^2 )(m_c^2 - m_q^2 -
s) + 2 s (m_q^2-m^2_{q^{'}})}{\lambda^{1/2} (s,s^\prime,q^2)
\lambda^{1/2}(m_c^2,m_q^2,s)} \le +1~,
\end{eqnarray}
where $\lambda(a,b,c)=a^2+b^2+c^2-2ab-2ac-2bc$ is the usual triangle function.

By the help of the Cutkosky rule, i.e., replacing the
propagators with the Dirac-delta functions:
\begin{eqnarray}\label{eq17}
\frac{1}{k^2-m^2} \rar -2i\pi \delta(k^2-m^2)~,
\end{eqnarray}

the spectral densities $\rho_i^{per}(s,s^{'},q^2)$ are found as:
\begin{eqnarray}  \label{eq18}
\rho_V &=&4 N_c~I_0(s,s^{\prime },q^2)~\{B_1(m_c-m_q)-B_2(m%
_{q^{\prime}}+m_q)-m_q\} ~,\nonumber \\ \nonumber\\
\rho_0 &=&-2 N_c~I_0(s,s^{\prime },q^2)~\{\Delta (m_q+m_{q^{'}})
-\Delta ^{\prime }(m_c-m_q)-4 A_1(m_c-%
m_q)
\nonumber \\
&&+2 m_q^2(m_c-m_q-m_{q^{'}})+m_q(2m_c m_{q^{'}}-u)%
\}~, \nonumber \\ \nonumber \\
\rho_1&=&2N_c~I_0(s,s^{\prime },q^2)\{B_1(m_c-%
3m_q)-B_2(m_q+m_{q^{'}})+2A_2(m_c-m_q)
\nonumber \\
&&+2A_3(m_c-m_q)-m_q\}~,\nonumber \\ \nonumber \\
\rho_2&=&2N_c~I_0(s,s^{\prime },q^2)\{2A_2(m_c-%
m_q)-2A_3(m_c-m_q)-B_1(m_c+m_q)
\nonumber \\
&&+B_2(m_q+m_{q^{'}})+m_q\}~.\nonumber \\ \nonumber \\
&&
\end{eqnarray}
where
\begin{eqnarray*}  \label{eq19}
I_{0}(s,s^{\prime},q^2)&=&\frac{1}{4\lambda^{1/2}(s,s^{\prime},q^2)},
\nonumber \\
\lambda(s,s^{\prime},q^2)&=&s^2+{s^{\prime}}^2+q^4-2sq^2-2s^{%
\prime}q^2-2ss^{\prime},  \nonumber \\
B_{1}&=&\frac{1}{\lambda(s,s^{\prime},q^2)}[2s^{\prime}\Delta-\Delta^{%
\prime}u],  \nonumber \\
B_{2}&=&\frac{1}{\lambda(s,s^{\prime},q^2)}[2s\Delta^{\prime}-\Delta
u],
\nonumber \\
A_{1}&=&\frac{1}{2\lambda(s,s^{\prime},q^2)}[{\Delta^{\prime}}%
^{2}s+\Delta^2s^{\prime}-
4m_{q^{'}}^2ss^{\prime}-\Delta\Delta^{\prime}u+m_{q}^2u^2],  \nonumber \\
A_{2}&=&\frac{1}{\lambda^{2}(s,s^{\prime},q^2)}[2{\Delta^{\prime}}%
^2ss^{\prime}+6\Delta^2{s^{\prime}}^2 -8m_{q}^2s{s^{\prime}}%
^2-6\Delta\Delta^{\prime}s^{\prime}u  \nonumber \\
&& +{\Delta^{\prime}}^2u^2 +2m_{q}^2s^{\prime}u^2],  \nonumber \\
A_{3}&=&\frac{1}{\lambda^{2}(s,s^{\prime},q^2)}[-3\Delta^2us^{\prime}-3{%
\Delta^{\prime}}^2us
+4m_{q}^2us^{\prime}s+4\Delta\Delta^{\prime}ss^{\prime}
\nonumber \\
&& +2\Delta\Delta^{\prime}u^2 -m_{q}^2u^3],  \nonumber \\
\end{eqnarray*}

where, $u=s+s^\prime-q^2$,
$\Delta=s+m_q^2-m_c^2$ , $\Delta'=s^\prime +m_q^2-m^2_{q^{'}}$ and  $N_c=3$ is the color factor.

 The corresponding non-perturbative part of the considered structures  are obtained as follows:
\begin{eqnarray} \label{eq20}
\Pi _{V}^{non-per}(p^{2},{p^{^{\prime }}}^{2},q^{2}) &=&<q\bar{q}>~\Bigg\{%
\frac{1\,}{2}\,{\frac{m_{{q}}m_{{q^{\prime }}}}{r{{r^{\prime }}}^{2}}}-\frac{%
1\,}{2}\,{\frac{m_{{q}}m_{{c}}}{{r}^{2}{r^{\prime
}}}}-{\frac{{m_{{q^{\prime
}}}}^{2}{m_{{q}}}^{2}}{r{{r^{\prime }}}^{3}}}+\frac{1\,}{2}\,{\frac{{m_{{%
q^{\prime }}}}^{2}{m_{{0}}}^{2}}{r{{r^{\prime }}}^{3}}}  \nonumber\\
&&-\frac{1\,}{2}\,{\frac{{m_{{c}}}^{2}{m_{{q}}}^{2}}{{r}^{2}{{r^{\prime }}}%
^{2}}}+\frac{1}{3}\,{\frac{{m_{{0}}}^{2}{m_{{c}}}^{2}}{{r}^{2}{{r^{\prime }}}%
^{2}}}-\frac{1\,}{2}\,{\frac{{m_{{q^{\prime }}}}^{2}{m_{{q}}}^{2}}{{r}^{2}{{%
r^{\prime }}}^{2}}}+\frac{1}{3}\,{\frac{{m_{{q^{\prime }}}}^{2}{m_{{0}}}^{2}%
}{{r}^{2}{{r^{\prime }}}^{2}}}+\frac{1\,}{2}{\frac{{q}^{2}{m_{{q}}}^{2}}{{r}%
^{2}{{r^{\prime }}}^{2}}}  \nonumber\\
&&-\frac{1}{3}\,{\frac{{m_{{0}}}^{2}{q}^{2}}{{r}^{2}{{r^{\prime }}}^{2}}}-{%
\frac{{m_{{c}}}^{2}{m_{{q}}}^{2}}{{r}^{3}{r^{\prime }}}}+\frac{1\,}{2}\,{%
\frac{{m_{{0}}}^{2}{m_{{c}}}^{2}}{{r}^{3}{r^{\prime }}}}+\frac{1}{6}\,{\frac{%
{m_{{0}}}^{2}m_{{c}}m_{{q^{\prime }}}}{{r}^{2}{{r^{\prime }}}^{2}}}+\frac{1}{%
3}\,{\frac{{m_{{0}}}^{2}}{{r}^{2}{r^{\prime }}}}\Bigg\}~,\nonumber\\
\end{eqnarray}
\begin{eqnarray} \label{eq21}
\Pi _{0}^{non-per}(p^{2},{p^{^{\prime }}}^{2},q^{2}) &=&<q\bar{q}>~\Bigg\{-%
\frac{1}{4}\,{\frac{m_{{q}}m_{{{q^{\prime }}}}}{r{r^{\prime }}}}-\frac{1}{4}%
\,{\frac{m_{{q}}{m_{{c}}}^{3}}{{r}^{2}{r^{\prime }}}}-\frac{1}{4}\,{\frac{m_{%
{q}}m_{{c}}}{r{r^{\prime }}}}+\frac{1}{4}\,{\frac{{m_{{0}}}^{2}{m_{{c}}}^{2}%
}{{r}^{2}{r^{\prime }}}}  \nonumber\\
&&+\frac{1}{4}\,{\frac{{m_{{0}}}^{2}{m_{{{q^{\prime }}}}}^{2}}{{r}^{2}{%
r^{\prime }}}}-\frac{1}{3}\,{\frac{{m_{{0}}}^{2}{q}^{2}}{{r}^{2}{r^{\prime }}%
}}+\frac{1}{6}\,{\frac{{m_{{0}}}^{2}{m_{{c}}}^{4}}{{r}^{2}{{r^{\prime }}}^{2}%
}}+\frac{1}{6}\,{\frac{{m_{{0}}}^{2}{m_{{{q^{\prime }}}}}^{4}}{{r}^{2}{{%
r^{\prime }}}^{2}}}+\frac{1}{6}\,{\frac{{m_{{0}}}^{2}{q}^{4}}{{r}^{2}{{%
r^{\prime }}}^{2}}}  \nonumber\\
&&+\frac{1}{6}\,{\frac{{m_{{0}}}^{2}{m_{{c}}}^{2}}{r{{r^{\prime }}}^{2}}}+%
\frac{1}{4}\,{\frac{{m_{{0}}}^{2}{m_{{{q^{\prime }}}}}^{2}}{r{{r^{\prime }}}%
^{2}}}-\frac{1}{6}\,{\frac{{m_{{0}}}^{2}{q}^{2}}{r{{r^{\prime }}}^{2}}}+%
\frac{1}{4}\,{\frac{m_{{q}}{m_{{{q^{\prime }}}}}^{3}}{r{{r^{\prime }}}^{2}}}-%
\frac{1\,}{2}{\frac{{m_{{{q^{\prime }}}}}^{4}{m_{{q}}}^{2}}{r{{r^{\prime }}}%
^{3}}} \nonumber \\
&&+\frac{1}{4}\,{\frac{{m_{{0}}}^{2}{m_{{{q^{\prime }}}}}^{4}}{r{{r^{\prime }%
}}^{3}}}-\frac{1}{4}\,{\frac{{m_{{c}}}^{4}{m_{{q}}}^{2}}{{r}^{2}{{r^{\prime }%
}}^{2}}}-\frac{1}{4}\,{\frac{{m_{{c}}}^{2}{m_{{q}}}^{2}}{{r}^{2}{r^{\prime }}%
}}-\frac{1}{4}\,{\frac{{m_{{{q^{\prime }}}}}^{2}{m_{{q}}}^{2}}{{r}^{2}{%
r^{\prime }}}}+\frac{1}{4}\,{\frac{{q}^{2}{m_{{q}}}^{2}}{{r}^{2}{r^{\prime }}%
}}  \nonumber\\
&&-\frac{1}{4}{\frac{{m_{{{q^{\prime }}}}}^{4}{m_{{q}}}^{2}}{{r}^{2}{{%
r^{\prime }}}^{2}}}-\frac{1}{4}\,{\frac{{m_{{{q^{\prime }}}}}^{2}{m_{{q}}}%
^{2}}{r{{r^{\prime }}}^{2}}}-\frac{1}{6}\,{\frac{{m_{{0}}}^{2}}{r{r^{\prime }%
}}}-\frac{3}{4}\,{\frac{{m_{{0}}}^{2}m_{{c}}m_{{{q^{\prime }}}}}{{r}^{2}{%
r^{\prime
}}}}+\frac{1}{6}\,{\frac{{m_{{0}}}^{2}{m_{{c}}}^{2}{m_{{{q^{\prime
}}}}}^{2}}{{r}^{2}{{r^{\prime }}}^{2}}}  \nonumber\\
&&-\frac{1}{3}\,{\frac{{m_{{0}}}^{2}{m_{{c}}}^{2}{q}^{2}}{{r}^{2}{{r^{\prime
}}}^{2}}}-\frac{1}{3}\,{\frac{{m_{{0}}}^{2}{m_{{{q^{\prime }}}}}^{2}{q}^{2}}{%
{r}^{2}{{r^{\prime }}}^{2}}}+\frac{1\,}{2}\,{\frac{m_{{q}}{m_{{c}}}^{2}m_{{{%
q^{\prime }}}}}{{r}^{2}{r^{\prime }}}}-\frac{1}{4}\,{\frac{m_{{q}}m_{{c}}{m_{%
{{q^{\prime }}}}}^{2}}{{r}^{2}{r^{\prime }}}} \nonumber \\
&&+\frac{1}{4}\,{\frac{m_{{q}}m_{{c}}{q}^{2}}{{r}^{2}{r^{\prime }}}}-\frac{1%
}{4}\,{\frac{m_{{q}}m_{{{q^{\prime }}}}{q}^{2}}{r{{r^{\prime }}}^{2}}}+\frac{%
1}{4}\,{\frac{m_{{q}}{m_{{c}}}^{2}m_{{{q^{\prime }}}}}{r{{r^{\prime }}}^{2}}}%
-\frac{1\,}{2}\,{\frac{m_{{q}}m_{{c}}{m_{{{q^{\prime
}}}}}^{2}}{r{{r^{\prime
}}}^{2}}}+\frac{1\,}{2}\,{\frac{{m_{{q}}}^{2}}{r{r^{\prime }}}} \nonumber \\
&&-\frac{1}{4}\,{\frac{{m_{{c}}}^{2}{m_{{q}}}^{2}}{r{{r^{\prime }}}^{2}}}-%
\frac{1}{4}{\frac{{q}^{4}{m_{{q}}}^{2}}{{r}^{2}{{r^{\prime }}}^{2}}}+\frac{1%
}{4}{\frac{{q}^{2}{m_{{q}}}^{2}}{r{{r^{\prime }}}^{2}}}-\frac{1\,}{2}\,{%
\frac{{m_{{c}}}^{4}{m_{{q}}}^{2}}{{r}^{3}{r^{\prime }}}}+\frac{1}{4}\,{\frac{%
{m_{{c}}}^{4}{m_{{0}}}^{2}}{{r}^{3}{r^{\prime }}}}  \nonumber\\
&&+\frac{1\,}{2}\,{\frac{m_{{c}}m_{{{q^{\prime }}}}{m_{{q}}}^{2}}{{r}^{2}{%
r^{\prime }}}}+\frac{1\,}{2}{\frac{m_{{c}}m_{{{q^{\prime }}}}{m_{{q}}}^{2}}{r%
{{r^{\prime }}}^{2}}}-\frac{1}{4}\,{\frac{m_{{c}}m_{{{q^{\prime }}}}{m_{{0}}}%
^{2}}{r{{r^{\prime }}}^{2}}}+{\frac{{m_{{c}}}^{3}m_{{{q^{\prime }}}}{m_{{q}}}%
^{2}}{{r}^{3}{r^{\prime }}}} \nonumber \\
&&-\frac{1\,}{2}{\frac{{m_{{c}}}^{3}m_{{{q^{\prime }}}}{m_{{0}}}^{2}}{{r}^{3}%
{r^{\prime }}}}-\frac{1\,}{2}\,{\frac{{m_{{c}}}^{2}{m_{{{q^{\prime }}}}}^{2}{%
m_{{q}}}^{2}}{{r}^{3}{r^{\prime }}}}+\frac{1}{4}\,{\frac{{m_{{c}}}^{2}{m_{{{%
q^{\prime }}}}}^{2}{m_{{0}}}^{2}}{{r}^{3}{r^{\prime }}}}+\frac{1\,}{2}\,{%
\frac{{m_{{c}}}^{2}{q}^{2}{m_{{q}}}^{2}}{{r}^{3}{r^{\prime }}}} \nonumber \\
&&-\frac{1}{4}\,{\frac{{m_{{0}}}^{2}{m_{{c}}}^{2}{q}^{2}}{{r}^{3}{r^{\prime }%
}}}+\frac{1\,}{2}{\frac{{m_{{c}}}^{2}{q}^{2}{m_{{q}}}^{2}}{{r}^{2}{{%
r^{\prime }}}^{2}}}-\frac{1\,}{2}\,{\frac{m_{{c}}m_{{{q^{\prime }}}}{q}^{2}{%
m_{{q}}}^{2}}{{r}^{2}{{r^{\prime }}}^{2}}}+\frac{1}{4}\,{\frac{m_{{c}}m_{{{%
q^{\prime }}}}{q}^{2}{m_{{0}}}^{2}}{{r}^{2}{{r^{\prime }}}^{2}}} \nonumber \\
&&+\frac{1\,}{2}\,{\frac{{m_{{{q^{\prime }}}}}^{2}{q}^{2}{m_{{q}}}^{2}}{{r}%
^{2}{{r^{\prime
}}}^{2}}}-\frac{1\,}{2}\,{\frac{{m_{{c}}}^{2}{m_{{{q^{\prime
}}}}}^{2}{m_{{q}}}^{2}}{r{{r^{\prime }}}^{3}}}+\frac{1}{4}\,{\frac{{m_{{c}}}%
^{2}{m_{{{q^{\prime }}}}}^{2}{m_{{0}}}^{2}}{r{{r^{\prime }}}^{3}}}+{\frac{m_{%
{c}}{m_{{{q^{\prime }}}}}^{3}{m_{{q}}}^{2}}{r{{r^{\prime }}}^{3}}} \nonumber \\
&&-\frac{1\,}{2}\,{\frac{m_{{c}}{m_{{{q^{\prime }}}}}^{3}{m_{{0}}}^{2}}{r{{%
r^{\prime }}}^{3}}}+\frac{1\,}{2}\,{\frac{{m_{{{q^{\prime }}}}}^{2}{q}^{2}{%
m_{{q}}}^{2}}{r{{r^{\prime }}}^{3}}}-\frac{1}{4}\,{\frac{{m_{{0}}}^{2}{m_{{{%
q^{\prime }}}}}^{2}{q}^{2}}{r{{r^{\prime }}}^{3}}}+\frac{1\,}{2}\,{\frac{{m_{%
{c}}}^{3}m_{{{q^{\prime }}}}{m_{{q}}}^{2}}{{r}^{2}{{r^{\prime }}}^{2}}} \nonumber \\
&&-\frac{1}{4}\,{\frac{{m_{{c}}}^{3}m_{{{q^{\prime }}}}{m_{{0}}}^{2}}{{r}^{2}%
{{r^{\prime }}}^{2}}}-\frac{1\,}{2}\,{\frac{{m_{{c}}}^{2}{m_{{{q^{\prime }}}}%
}^{2}{m_{{q}}}^{2}}{{r}^{2}{{r^{\prime }}}^{2}}}+\frac{1\,}{2}\,{\frac{m_{{c}%
}{m_{{{q^{\prime }}}}}^{3}{m_{{q}}}^{2}}{{r}^{2}{{r^{\prime }}}^{2}}}-\frac{1%
}{4}\,{\frac{m_{{c}}{m_{{{q^{\prime }}}}}^{3}{m_{{0}}}^{2}}{{r}^{2}{{%
r^{\prime }}}^{2}}}\Bigg\}~,\nonumber\\
\end{eqnarray}
\begin{eqnarray} \label{eq22}
\Pi _{1}^{non-per}(p^{2},{p^{^{\prime }}}^{2},q^{2}) &=&<q\bar{q}>~\Bigg\{-%
\frac{1}{4}\,{\frac{m_{{q}}m_{{q^{\prime }}}}{r{{r^{\prime }}}^{2}}}+\frac{1%
}{4}\,{\frac{m_{{q}}m_{{c}}}{{r}^{2}{r^{\prime }}}}+\frac{1\,}{2}\,{\frac{{%
m_{{q^{\prime }}}}^{2}{m_{{q}}}^{2}}{r{{r^{\prime }}}^{3}}}-\frac{1}{4}\,{%
\frac{{m_{{q^{\prime }}}}^{2}{m_{{0}}}^{2}}{r{{r^{\prime }}}^{3}}}  \nonumber\\
&&+\frac{1}{4}\,{\frac{{m_{{c}}}^{2}{m_{{q}}}^{2}}{{r}^{2}{{r^{\prime }}}^{2}%
}}-\frac{1}{6}\,{\frac{{m_{{0}}}^{2}{m_{{c}}}^{2}}{{r}^{2}{{r^{\prime }}}^{2}%
}}+\frac{1}{4}\,{\frac{{m_{{q^{\prime }}}}^{2}{m_{{q}}}^{2}}{{r}^{2}{{%
r^{\prime }}}^{2}}}-\frac{1}{6}\,{\frac{{m_{{q^{\prime }}}}^{2}{m_{{0}}}^{2}%
}{{r}^{2}{{r^{\prime }}}^{2}}}-\frac{1}{4}\,{\frac{{q}^{2}{m_{{q}}}^{2}}{{r}%
^{2}{{r^{\prime }}}^{2}}}  \nonumber\\
&&+\frac{1}{6}\,{\frac{{m_{{0}}}^{2}{q}^{2}}{{r}^{2}{{r^{\prime }}}^{2}}}+%
\frac{1\,}{2}\,{\frac{{m_{{c}}}^{2}{m_{{q}}}^{2}}{{r}^{3}{r^{\prime }}}}-%
\frac{1}{4}\,{\frac{{m_{{0}}}^{2}{m_{{c}}}^{2}}{{r}^{3}{r^{\prime }}}}-\frac{%
1\,}{2}\,{\frac{{m_{{q}}}^{2}}{{r}^{2}{r^{\prime }}}}+\frac{1}{6}\,{\frac{{%
m_{{0}}}^{2}}{{r}^{2}{r^{\prime }}}}  \nonumber\\
&&-\frac{1\,}{12}\,{\frac{{m_{{0}}}^{2}m_{{c}}m_{{q^{\prime }}}}{{r}^{2}{{%
r^{\prime }}}^{2}}}\Bigg\}~,
\end{eqnarray}
\begin{eqnarray} \label{eq23}
\Pi _{2}^{non-per}(p^{2},{p^{^{\prime }}}^{2},q^{2}) &=&<q\bar{q}>~\Bigg\{%
\frac{1}{4}\,{\frac{m_{{q}}m_{{q^{\prime }}}}{r{{r^{\prime }}}^{2}}}-\frac{1%
}{4}\,{\frac{m_{{q}}m_{{c}}}{{r}^{2}{r^{\prime }}}}-\frac{1\,}{2}\,{\frac{{%
m_{{q^{\prime }}}}^{2}{m_{{q}}}^{2}}{r{{r^{\prime }}}^{3}}}+\frac{1}{4}\,{%
\frac{{m_{{q^{\prime }}}}^{2}{m_{{0}}}^{2}}{r{{r^{\prime }}}^{3}}}  \nonumber\\
&&-\frac{1}{4}{\frac{{m_{{c}}}^{2}{m_{{q}}}^{2}}{{r}^{2}{{r^{\prime }}}^{2}}}%
+\frac{1}{6}{\frac{{m_{{0}}}^{2}{m_{{c}}}^{2}}{{r}^{2}{{r^{\prime }}}^{2}}}-%
\frac{1}{4}{\frac{{m_{{q^{\prime }}}}^{2}{m_{{q}}}^{2}}{{r}^{2}{{r^{\prime }}%
}^{2}}}+\frac{1}{6}\,{\frac{{m_{{q^{\prime }}}}^{2}{m_{{0}}}^{2}}{{r}^{2}{{%
r^{\prime }}}^{2}}}+\frac{1}{4}\,{\frac{{q}^{2}{m_{{q}}}^{2}}{{r}^{2}{{%
r^{\prime }}}^{2}}} \nonumber \\
&&-\frac{1}{6}\,{\frac{{m_{{0}}}^{2}{q}^{2}}{{r}^{2}{{r^{\prime }}}^{2}}}-%
\frac{1\,}{2}\,{\frac{{m_{{c}}}^{2}{m_{{q}}}^{2}}{{r}^{3}{r^{\prime }}}}+%
\frac{1}{4}\,{\frac{{m_{{0}}}^{2}{m_{{c}}}^{2}}{{r}^{3}{r^{\prime }}}}-\frac{%
1\,}{2}\,{\frac{{m_{{q}}}^{2}}{{r}^{2}{r^{\prime }}}}+\frac{1\,}{2}\,{\frac{{%
m_{{0}}}^{2}}{{r}^{2}{r^{\prime }}}}  \nonumber\\
&&+\frac{1\,}{12}\,{\frac{{m_{{0}}}^{2}m_{{c}}m_{{q^{\prime }}}}{{r}^{2}{{%
r^{\prime }}}^{2}}}\Bigg\}~.
\end{eqnarray}
where $r=p^2-m_c^2$, $r^{\prime}={p^{\prime}}^2-m_{q^{\prime}}^2$.

 Equating two representations of the correlation
function and applying the double Borel transformation using
\begin{eqnarray}  \label{eq24}
{\cal{B}}_{p^2}(M_1^2)(\frac{1}{p^2-m^2_c})^m=\frac{(-1)^m}{\Gamma(m)}
\frac{e^{-\frac{m_c^2}{M_1^2}}}{(M_1^2)^m}, \nonumber \\
{\cal{B}}_{{p^{'}}^2}(M_2^2)(\frac{1}{{p^{'}}^2-m^2_{q^{'}}})^n=\frac{(-1)^n}{\Gamma(n)}
\frac{e^{-\frac{m_{q'}^2}{M_2^2}}}{(M_2^2)^n},
\end{eqnarray}
 the sum rules for
the form factors $F_i^{D_{(s)}\to K_1}$ are obtained as:
\begin{eqnarray}\label{eq25}
F_{i}^{D_{(s)}\to K_1} \es  -\frac{(m_c+m_q)}{f_{D_q} m_{D_q}^2
f_{K_1} m_{K_1}} e^\frac{m_{D_q}^2}{M_1^2}
e^\frac{m_{K_1}^2}{M_2^2} \Bigg\{-\frac{1}{4 \pi^2}
\int_{m_c^2}^{s_0^\prime} ds^\prime \int_{s_L}^{s_0} ds\rho_{i}
(s,s^\prime,q^2) e^\frac{-s}{M_1^2}
e^\frac{-s^{'}}{M_2^2} \nnb \\
&+& M^{2}_{1}~M^{2}_{2}~~
{\cal{B}}_{p^2}(M_1^2)~{\cal{B}}_{{p^{'}}2}(M_2^2)~[\Pi_{i}^{non-per}(p^2,{p^{'}}^2,q^2)]
\Bigg\},
\end{eqnarray}
where $i=V,0,1$ and $2$, $s_0$ and $s_0^{'}$ are the continuum
thresholds in pseudoscalar $D_q$ and axial-vector $K_1$ channels,
respectively and the lower limit in the  integration over $s$ is as
follows:
\begin{eqnarray}\label{eq26}
s_L \es \frac{(m_q^2+q^2-m_c^2-s^\prime) (m_c^2 s^\prime -m_q^2
q^2)} {(m_c^2-q^2) (m_q^2-s^\prime)}~.
\end{eqnarray}
In
Eq.~(\ref{eq25}),  to subtract the contributions of the
higher states and the continuum the quark-hadron duality assumption
is also
 used, i.e., it is assumed
that
\begin{eqnarray}
\rho^{higher states}(s,s') = \rho^{OPE}(s,s') \theta(s-s_0)
\theta(s-s'_0).
\end{eqnarray}
 Here, we should stress that in the three-point sum rules with double dispersion relation,
 the subtraction of the continuum states and the quark-hadron
duality is highly nontrivial. For $q^2>0$ values, their may be an
inconsistency between double dispersion integrals in Eq.
(\ref{eq25}) and corresponding coefficients  of the structures in
the Feynman amplitudes in the bare-loop diagram. In this case, the
double spectral density receives contributions beyond the
contributions coming from the Landau-type singularities.
 This problem has been widely discussed   in \cite{Ball1}.
Here, we neglect such contributions since  with the above continuum subtraction
and the selecting integration region the contribution of the non-Landau singularities is very small comparing the Landau type singularity contributions.

Now, as we mentioned in the introduction section, the $F_i^{D_{q}\to
K_{1A(B)}}$ form factors are obtained from the above equation by
replacing  $f_{K_1}$ by the G-parity conserving decay constant
$f_{K_{1A}}$  and G-parity violating decay constant
$f_{K_{1B}}=f_{K_{1B^{\perp}}}(1~~GeV)a_{0}^{\parallel,K_{1B}}$ and
$m_{K_1}$ with $m_{K_{1A(B)}}$, i.e.,
\begin{eqnarray}\label{eq27}
F_{i}^{D_{(s)}\to K_{1A(B)}} \es  -\frac{(m_c+m_q)}{f_{D_q}
m_{D_q}^2 f_{K_{1A(B)}} m_{K_{1A(B)}}} e^\frac{m_{D_q}^2}{M_1^2}
e^\frac{m_{K_{1A(B)}}^2}{M_2^2} \Bigg\{-\frac{1}{4 \pi^2}
\int_{m_c^2}^{s_0^\prime} ds^\prime \int_{s_L}^{s_0}ds \rho_{i}
(s,s^\prime,q^2) e^\frac{-s}{M_1^2}
e^\frac{-s^{'}}{M_2^2} \nnb \\
&+& M^{2}_{1}~M^{2}_{2}~~
{\cal{B}}_{p^2}(M_1^2)~{\cal{B}}_{{p^{'}}2}(M_2^2)~[\Pi_{i}^{non-per}(p^2,{p^{'}}^2,q^2)]
\Bigg\}.
\end{eqnarray}
 Also, using  Eqs. (\ref{eq1}, \ref{eq4}, \ref{eq5}, \ref{eq6}), the form factors of the $f_i^{D_{q}\to K_{1}[1270(1400)]}$ are found as follows:
\begin{eqnarray}\label{eq29}
f_0^{D_{q}\to
K_1(1270)}=(\frac{m_{D_q}+m_{K_{1A}}}{m_{D_q}+m_{K_{1}}})~f_0^{D_{q}\to
K_{1A}}~sin\theta_{K_1}+(\frac{m_{D_q}+m_{K_{1B}}}{m_{D_q}+m_{K_{1}}})~f_0^{D_{q}\to
K_{1B}}~cos\theta_{K_1}~,\nonumber\\
f_{1,2,V}^{D_{q}\to
K_1(1270)}=(\frac{m_{D_q}+m_{K_{1}}}{m_{D_q}+m_{K_{1A}}})~f_{1,2,V}^{D_{q}\to
K_{1A}}~sin\theta_{K_1}+(\frac{m_{D_q}+m_{K_{1}}}{m_{D_q}+m_{K_{1B}}})~f_{1,2,V}^{D_{q}\to
K_{1B}}~cos\theta_{K_1}~,\nonumber\\
f_0^{D_{q}\to
K_1(1400)}=(\frac{m_{D_q}+m_{K_{1A}}}{m_{D_q}+m_{K_{1}}})~f_0^{D_{q}\to
K_{1A}}~cos\theta_{K_1}-(\frac{m_{D_q}+m_{K_{1B}}}{m_{D_q}+m_{K_{1}}})~f_0^{D_{q}\to
K_{1B}}~sin\theta_{K_1}~,\nonumber\\
f_{1,2,V}^{D_{q}\to
K_1(1400)}=(\frac{m_{D_q}+m_{K_{1}}}{m_{D_q}+m_{K_{1A}}})~f_{1,2,V}^{D_{q}\to
K_{1A}}~cos\theta_{K_1}-(\frac{m_{D_q}+m_{K_{1}}}{m_{D_q}+m_{K_{1B}}})~f_{1,2,V}^{D_{q}\to
K_{1B}}~sin\theta_{K_1}~.\nonumber\\
\end{eqnarray}

\section{Decay amplitudes and decay widths }
\section*{semileptonic}
Using the amplitude in Eq. (\ref{eq3}) and  definitions of  the form factors, the differential
decay widths  for the process $D_q \rightarrow
K_1 \ell \nu$ are found as follows:
\begin{eqnarray}\label{eq30}
\frac{d\Gamma _{\pm }(D_{q}\rightarrow K_1 \ell \nu)}{dq^2}
&=&\frac{G_F^{2}\left| V_{cq^{'}}\right| ^{2}}{192\pi
^{3}m^3_{D_q}}~q^{2}\lambda
^{1/2}(m^2_{D_q},m^2_{K_1},q^{2})\left|
H_{\pm }\right| ^{2}~,  \nonumber \\
&&  \nonumber \\
\frac{d\Gamma _{0}(D_{q}\rightarrow K_1 \ell \nu)}{dq^2}
&=&\frac{G_F^{2}\left| V_{cq^{'}}\right| ^{2}}{192\pi
^{3}m^3_{D_q}}~ q^{2}\lambda
^{1/2}(m^2_{D_q},m^2_{K_1},q^{2})\left| H_{0}\right| ^{2}~,
\end{eqnarray}
where,
\begin{eqnarray*}\label{eq31}
H_{\pm }(q^{2})
&=&(m_{D_q}+m_{K_1})f_{0}(q^{2})\mp \frac{%
\lambda ^{1/2}(m^2_{D_q},m^2_{K_1},q^{2})}{%
m_{D_q}+m_{K_1}}f_{V}(q^{2})~, \\
\\
H_{0}(q^{2}) &=&\frac{1}{2m_{K_1}\sqrt{q^2}}\Bigg[%
(m^2_{D_q}-m^2_{K_1}-q^{2})(m_{D_q}+m_{K_1}%
)f_{0}(q^{2}) -\frac{\lambda
(m^2_{D_q},m^2_{K_1},q^{2})}{%
m_{D_q}+m_{K_1}}f_{1}(q^{2})\Bigg]~.
\end{eqnarray*}
The $\pm, 0$ in the above relations   belong  to the $K_1$
helicities. The total differential decay width can be written as
\begin{eqnarray}  \label{eq50}
\frac{d\Gamma _{tot}(D_{q}\rightarrow K_1 \ell
\nu)}{dq^2}=\frac{d\Gamma _{L }(D_{q}\rightarrow K_1 \ell
\nu)}{dq^2}+\frac{d\Gamma _{T}(D_{q}\rightarrow K_1 \ell
\nu)}{dq^2},
\end{eqnarray}
where,
\begin{eqnarray}  \label{eq51}
\frac{d\Gamma _{L}(D_{q}\rightarrow K_1 \ell \nu)}{dq^2}
&=&\frac{d\Gamma _{0 }(D_{q}\rightarrow K_1 \ell \nu)}{dq^2},\nonumber\\
\frac{d\Gamma _{T}(D_{q}\rightarrow K_1 \ell \nu)}{dq^2}
&=&\frac{d\Gamma _{+ }(D_{q}\rightarrow K_1 \ell
\nu)}{dq^2}+\frac{d\Gamma _{- }(D_{q}\rightarrow K_1 \ell
\nu)}{dq^2},
\end{eqnarray}
and $\frac{d\Gamma _{L}}{dq^2}$  ($\frac{d\Gamma _{T}}{dq^2}$) is
the longitudinal  (transverse) component of the differential decay
width.
\section*{nonleptonic}
In this part, we study the  decay amplitude and decay width for the nonleptonic  $D\rightarrow
K_1 \pi$ decay. The effective Hamiltonian for this decay at the
quark level is given by (see for example \cite{azizi1} and references therein):
\begin{eqnarray}\label{eq32}
H_{eff} &=&{G_F\over \sqrt{2}} \left \{ V_{cs} V^*_{ud}(C_1 O_1
+C_2 O_2 ) \right \}.
\end{eqnarray}
Here $O_1$ and $O_2$  are quark  operators and they are given as:
\begin{equation}\begin{array}{llllll}
O_{1}&=&(\bar{s_{i}}c_{i})_{V-A}(\bar{u}_{j}d_{j})_{V-A},  &
O_{2}&=&(\bar{s}_{i}c_{j})_{V-A}(\bar{u}_{j}d_{i})_{V-A},
\end{array}\end{equation}
where $(\bar q_{1}q_{2})_{V\pm A}=\bar
q_{1}\gamma^{\mu}(1\pm\gamma_{5})q_{2}$.

The Wilson coefficients $C_1$ and $C_2$ have been calculated in
different schemes \cite{Buchalla}. In the present work, we will use
$C_1(m_c)=1.263$ and $C_2(m_c)=-0.513$ obtained at the leading
order in renormalization group improved perturbation theory at
$\mu\simeq1.3~GeV$ \cite{Colangelo1}.

Now, we calculate the amplitude $\mathcal{A}$ for $D\rightarrow K_1
\pi$ decay. Using the factorization method and definition of the
related matrix elements in terms of the form factors $f^{D\to
K_1}_{V},~f^{D\to K_1}_{0},~f^{D\to K_1}_{1}$ and $f^{D\to K_1}_{2}$
in Eqs. (\ref{eq4}-\ref{eq6}), we obtain this amplitude as follows:
\begin{eqnarray}\label{eq33}
\mathcal{A}^{D\rightarrow K_1 \pi} &=&\frac{G_{F}}{\sqrt{2}}%
~\{V_{cs}V_{ud}^{\ast }~a_{1}\}~f_{\pi}~(\varepsilon.
p)~[F^{D\rightarrow K_1 \pi}(m_{\pi}^{2})],
\end{eqnarray}
where,
\begin{eqnarray}\label{eq34}
F^{D\rightarrow K_1 \pi}(m_{\pi}^{2})
&=&[(m_{D}+m_{K_1})f_{0}(m_{\pi}^{2})
-(m_{D}-m_{K_1})f_{1}(m_{\pi}^{2})-\frac{f_{2}(m_{\pi}^{2})}{(m_{D}+m_{K_1})}m_{\pi}^{2}]~.
\end{eqnarray}
The $\varepsilon$ stands for polarization of $K_1$, $p$ is
four momentum of $D$, $f_{\pi}$ is the  pion decay constant,
$a_1=C_1+\frac{1}{N_c} C_2$ and $N_c$ is the  number of colors in QCD.

Now, we can calculate the decay width for $D\rightarrow K_1 \pi$ decay.
The explicit expression for decay width is given as follow:
\begin{eqnarray}\label{eq35}
\Gamma (D\rightarrow K_1 \pi)&=&\frac{G_{F}^{2}}{128~\pi
m_{D}^{3}m_{K_1}^{2}}|V_{cs}|
^{2}|V_{ud }|^{2}~a_{1}^{2}~f_{\pi}^{2}\nonumber \\
&&\lambda(m_{D}^{2},m_{K_1}^{2},m_{\pi}^{2})^\frac{3}{2}~
[F^{D\rightarrow K_1 \pi}(m_\pi^{2})]^{2}.
\end{eqnarray}

\section{Numerical analysis}
From the sum rules expressions of the form factors, it is clear
that the main input parameters entering the expressions are
condensates, elements of the CKM matrix $V_{cq^{'}}$, leptonic
decay constants $f_{D_q}$,  $f_{K_1A}$ and $f_{K_{1B^{\perp}}}$,
Borel parameters $M_{1}^2$ and $M_{2}^2$ as well as the continuum
thresholds $s_{0}$ and $s'_{0}$. We choose the values of the
condensates (at a fixed renormalization scale of about $1~GeV$),
leptonic decay constants , CKM matrix elements, quark and meson
masses as: $<u\bar{u}>=<d\bar{d}>=-(0.240\pm 0.010~GeV)^3$,
$<s\bar{s}>=(0.8\pm 0.2) <u\bar{u}>$, $m_0^2=0.8\pm 0.2~GeV^2$
\cite{Ioffe}, $\mid V_{cs}\mid=0.957\pm0.110$, $\mid V_{cd}\mid=0.230\pm0.011$
\cite{Yao}, $f_{D^0}=f_{D^\pm}=0.222\pm0.016 ~GeV$ \cite{Artuso2},
$f_{D_s}=0.274\pm0.013~GeV$ \cite{Artuso1},
$f_{K_{1A}}=0.250\pm0.013~GeV$,
$f_{K_{1B^{\perp}}}=0.190\pm0.010~GeV$ \cite{Lee},
$m_{u}(1~GeV)=(1.5-3.3)~MeV$, $m_{d}(1~GeV)=(3.5-6)~MeV$,
$m_{s}(1~GeV)=(104^{+26}_{-34})~MeV$,
$m_{c}=1.27^{+0.07}_{-0.11}~GeV$, $m_{D^0}=1.864~GeV$,
$m_{D^\pm}=1.869~GeV$, $m_{D_s}=1.968~GeV$,
$m_{K_1}(1270)=1.27~GeV$, $m_{K_1}(1400)=1.40~GeV$\cite{Yao},
$m_{K_{1A}}=1.31\pm0.06~GeV$, $m_{K_{1B}}=1.34\pm0.08~GeV$ and
$a_{0}^{\parallel,K_{1B}}=-0.19\pm0.07$ \cite{Lee}.

The sum rules for the form factors  contain also four auxiliary
parameters: Borel mass squares $M_{1}^2$ and $M_{2}^2$ and continuum
thresholds $s_{0}$ and $s'_{0}$. These are not physical quantities,
so the  form factors as physical quantities should be independent of
them. The parameters $s_0$ and $s_0^\prime$, which are the continuum
thresholds of $D_q$ and $K_1$ mesons, respectively, are determined
from the condition that guarantees the sum rules to practically be
stable  in the allowed regions for $M_1^2$ and $M_2^2$. The values
of the continuum thresholds calculated from the two--point QCD sum
rules are taken to be $s_0=(6-8)~GeV^2$ and
$s_0^\prime=(4-6)~GeV^2$. The working regions for $M_1^2$ and
$M_2^2$ are determined requiring that not only the contributions of
the higher states and continuum are small, but the contributions of
the operators with higher dimensions are also small. Both conditions
are satisfied in the  regions $4~GeV^2 \le M_1^2 \le 10~GeV^2$ and
$3~GeV^2 \le M_2^2 \le 8~GeV^2$.

The values of the form factors at $q^2=0$ are  shown in Tables 1 and
 2. Note that, the values of the  $f_i(0)$ for $D^0\to K_1^\pm \ell\nu$
and $D^\pm\to K_1^0 \ell\nu$ are approximately equal, so the values
in Table. 1 refer to both decays.
\begin{table}[h]\label{tab:3}
\vspace{0cm} \centering
\begin{tabular}{c|cccc||c|cccc}
\cline{2-5}\cline{7-10} $\theta_{K_1}^\circ$ &37&58&-37&-58&$\theta_{K_1}^\circ$ &37&58&-37&-58\\
\cline{1-10}\lower0.35cm \hbox{{\vrule width 0pt height 1.0cm }}
$f_{V}^{D\to K_1(1270)}$ &3.19&1.82 &4.00 &2.95 &$f_{V}^{D\to K_1(1400)}$&-3.37 &-4.34&2.27&3.60 \\
\cline{1-10}\lower0.35cm \hbox{{\vrule width 0pt height 1.0cm }}
$f_{0}^{D\to K_1(1270)}$ &-0.74&-0.42 &-0.93 &-0.68 &$f_{0}^{D\to K_1(1400)}$&0.72 &0.92&-0.49&-0.77 \\
\cline{1-10}\lower0.35cm \hbox{{\vrule width 0pt height 1.0cm }}
$f_{1}^{D\to K_1(1270)}$ &0.34& 0.19 &0.44 &0.34 &$f_{1}^{D\to K_1(1400)}$&-0.38 &-0.49&0.23&0.38 \\
\cline{1-10}\lower0.35cm \hbox{{\vrule width 0pt height 1.0cm }}
$f_{2}^{D\to K_1(1270)}$ &2.56&1.46 &3.24 &2.36 &$f_{2}^{D\to K_1(1400)}$&-2.70 &-3.49&1.82&2.90 \\
\cline{1-10}
\end{tabular}
\vspace{0.5cm} \caption{The $q^2=0$ values of the form factors of
the $D\to K_1 \ell \nu$ decay for $M_1^2=8~GeV^2$, $M_2^2=6~GeV^2$
at  different values of $\theta_{K_1}$.}
\end{table}

\begin{table}[h]\label{tab:3}
\vspace{0cm}\centering
\begin{tabular}{c|cccc||c|cccc}
\cline{2-5}\cline{7-10} $\theta_{K_1}^\circ$ &37&58&-37&-58&$\theta_{K_1}^\circ$ &37&58&-37&-58\\
\cline{1-10}\lower0.35cm \hbox{{\vrule width 0pt height 1.0cm }}
$f_{V}^{D_s^+\to K_1^0(1270)}$ &3.90&2.22 &4.86 &3.58 &$f_{V}^{D_s^+\to K_1^0(1400)}$&-4.09 &-5.27&2.76&4.40\\
\cline{1-10}\lower0.35cm \hbox{{\vrule width 0pt height 1.0cm }}
$f_{0}^{D_s^+\to K_1^0(1270)}$ &-1.15&-0.65 &-1.44 &-1.07&$f_{0}^{D_s^+\to K_1^0(1400)}$&1.12 &1.44&-0.76&-1.20 \\
\cline{1-10}\lower0.35cm \hbox{{\vrule width 0pt height 1.0cm }}
$f_{1}^{D_s^+\to K_1^0(1270)}$ &-0.54&-0.31 &-0.66 &-0.50 &$f_{1}^{D_s^+\to K_1^0(1400)}$&0.57 &0.73&-0.39&-0.61 \\
\cline{1-10}\lower0.35cm \hbox{{\vrule width 0pt height 1.0cm }}
$f_{2}^{D_s^+\to K_1^0(1270)}$ &5.89&3.36 &7.33 &5.40 &$f_{2}^{D_s^+\to K_1^0(1400)}$&-6.19 &-7.97&4.18&6.64 \\
\cline{1-10}
\end{tabular}
\vspace{0.50cm} \caption{The $q^2=0$ values of the form factors of
the $D_s\to K_1 \ell \nu$ decay for $M_1^2=8~GeV^2$, $M_2^2=6~GeV^2$
at different values of $\theta_{K_1}$.}
\end{table}

The dependence of the $f_i^{D_{q}\to K_1}(0)$ on $\theta_{K_1}$ at
$q^2=0$ is depicted in Figs. 5-8,~in the interval
$-58^\circ\leq\theta_{K_1}\leq58^\circ$. In Figs. 6 and 8, as it is
seen, all of the form factors contact at one point. Also each form
factor in Figs. 5 and 7, has one extremum point. These extrema as
well as the contact points have been specified in Figs. 5-8. It is
interesting that in the $D_{q}\to K_1(1270) \ell\nu$ and $D_{q}\to
K_1(1400) \ell\nu$ cases, the extrema and contact points of the form
factors  are nearly at $-8^\circ$. The sum rules for  the form
factors are truncated at about $q^2=0.15~ GeV^2$ and $q^2=0.25~
GeV^2$ for $q=u(d)$ and s cases of the  $K_1(1270)$, respectively.
These points for $K_1(1400)$ state are $q^2=0.22~ GeV^2$ and
$q^2=0.32~ GeV^2$ for u(d) and s cases, respectively. To  extend the
results to the full physical region, i.e., $0 \leq q^2 \leq
(m_{D_{q}}-m_{K_1})^2~ GeV^2$, we look for a parametrization such
that: 1) this parametrization coincides well with the sum rules
predictions below the points at which the form factors are truncated
and 2) the parametrization provides an extrapolation to $q^2>$ the
truncated points, which is consistent with the expected analytical
properties of the form factors and reproduces the lowest-lying
resonance (pole). This  resonance in the $D_q$ channel is
$D^*(J^P=1^-)$ state. Following  references \cite{rev21,rev22},
which describe this point in details, we choose the following
theoretically more reliable fit parametrization:
\newline
\begin{equation}  \label{eq36}
f_{i}(q^2)=\frac{a}{1- \frac{q^2}{m^{2}_{D^*}}}+ \frac{b}{1-
\frac{q^2}{m^{2}_{fit}}}.
\end{equation}
The values of the parameters $%
a,~b$ and $m_{fit}$ are given in  Tables 3-6 at
  different values of the mixing angle $\theta_{K_1}$. From this
  parametrization, we see that the $m_{D^*}$ pole exist outside the
  allowed physical region and related to that, one can calculate the
  hadronic parameters such as the coupling constant $g_{DD^*K_1}$
  (see \cite{rev23,rev24}).
\begin{table}[h]\label{tab:3}
\centering
\begin{tabular}{c|ccc||c|ccc}
\cline{2-4}\cline{6-8}& $a$ &$ b $& $m_{fit}$ & & $ a $& $b$ & $m_{fit}$\\
\cline{1-8}\lower0.35cm \hbox{{\vrule width 0pt height 1.0cm }}
$f_{V}^{D\to K_1(1270)}(q^2)$ &3.83&-0.64 &1.25&$f_{V}^{D\to K_1(1400)}(q^2)$&-5.94 &2.57&1.25\\
\cline{1-8}\lower0.35cm \hbox{{\vrule width 0pt height 1.0cm }}
$f_{0}^{D\to K_1(1270)}(q^2)$ &-2.05&1.31 &1.36&$f_{0}^{D\to K_1(1400)}(q^2)$&2.04 &-1.32&1.36\\
\cline{1-8}\lower0.35cm \hbox{{\vrule width 0pt height 1.0cm }}
$f_{1}^{D\to K_1(1270)}(q^2)$ &0.46&-0.12 &1.27&$f_{1}^{D\to K_1(1400)}(q^2)$&-0.59&0.21&1.27\\
\cline{1-8}\lower0.35cm \hbox{{\vrule width 0pt height 1.0cm }}
$f_{2}^{D\to K_1(1270)}(q^2)$ &2.97&-0.41 &1.29&$f_{2}^{D\to K_1(1400)}(q^2)$&-3.14 &0.44&1.29\\
\cline{1-8}\lower0.35cm \hbox{{\vrule width 0pt height 1.0cm }}
$f_{V}^{D_s^+\to K_1^0(1270)}(q^2)$ &4.08&-0.18 &1.28&$f_{V}^{D_s^+\to K_1^0(1400)}(q^2)$&-7.87 &3.78&1.28\\
\cline{1-8}\lower0.35cm \hbox{{\vrule width 0pt height 1.0cm }}
$f_{0}^{D_s^+\to K_1^0(1270)}(q^2)$ &-3.56&2.41 &1.51&$f_{0}^{D_s^+\to K_1^0(1400)}(q^2)$&3.06 &-1.94&1.51\\
\cline{1-8}\lower0.35cm \hbox{{\vrule width 0pt height 1.0cm }}
$f_{1}^{D_s^+\to K_1^0(1270)}(q^2)$ &-0.70&0.16&1.31&$f_{1}^{D_s^+\to K_1^0(1400)}(q^2)$&0.58 &-0.01&1.31\\
\cline{1-8}\lower0.35cm \hbox{{\vrule width 0pt height 1.0cm }}
$f_{2}^{D_s^+\to K_1^0(1270)}(q^2)$ &7.12&-1.23 &1.35&$f_{2}^{D_s^+\to K_1^0(1400)}(q^2)$&-5.32 &-0.87&1.35\\
\cline{1-8}
\end{tabular}
\vspace{0.01cm} \caption{Parameters appearing in the fit function
for the form factors of the $D_{q}\to K_1(1270)\ell\nu$ and
$D_{q}\to K_1(1400)\ell\nu$ decays at $M_1^2=8~GeV^2$,
$M_2^2=6~GeV^2$ and $\theta_{K_1}=37^\circ$.}
\end{table}
\begin{table}[h]\label{tab:3}
\centering
\begin{tabular}{c|ccc||c|ccc}
\cline{2-4}\cline{6-8}&$a$&$b$&$m_{fit}$&&$a$&$b$&$m_{fit}$\\
\cline{1-8}\lower0.35cm \hbox{{\vrule width 0pt height 1.0cm }}
$f_{V}^{D\to K_1(1270)}(q^2)$ &2.12&-0.30 &1.27&$f_{V}^{D\to K_1(1400)}(q^2)$&-7.44 &3.10&1.27\\
\cline{1-8}\lower0.35cm \hbox{{\vrule width 0pt height 1.0cm }}
$f_{0}^{D\to K_1(1270)}(q^2)$ &-1.52&1.10 &1.37&$f_{0}^{D\to K_1(1400)}(q^2)$&2.70 &-1.78&1.37\\
\cline{1-8}\lower0.35cm \hbox{{\vrule width 0pt height 1.0cm }}
$f_{1}^{D\to K_1(1270)}(q^2)$ &0.27&-0.08 &1.29&$f_{1}^{D\to K_1(1400)}(q^2)$&-0.75&0.26&1.29\\
\cline{1-8}\lower0.35cm \hbox{{\vrule width 0pt height 1.0cm }}
$f_{2}^{D\to K_1(1270)}(q^2)$ &1.68&-0.22 &1.31&$f_{2}^{D\to K_1(1400)}(q^2)$&-4.00 &0.51&1.31\\
\cline{1-8}\lower0.35cm \hbox{{\vrule width 0pt height 1.0cm }}
$f_{V}^{D_s^+\to K_1^0(1270)}(q^2)$ &1.29&0.93 &1.30&$f_{V}^{D_s^+\to K_1^0(1400)}(q^2)$&-9.18&3.91&1.30\\
\cline{1-8}\lower0.35cm \hbox{{\vrule width 0pt height 1.0cm }}
$f_{0}^{D_s^+\to K_1^0(1270)}(q^2)$ &-2.14&1.49 &1.53&$f_{0}^{D_s^+\to K_1^0(1400)}(q^2)$&4.03 &-2.59&1.53\\
\cline{1-8}\lower0.35cm \hbox{{\vrule width 0pt height 1.0cm }}
$f_{1}^{D_s^+\to K_1^0(1270)}(q^2)$ &-0.45&0.14 &1.32&$f_{1}^{D_s^+\to K_1^0(1400)}(q^2)$&0.79 &-0.06&1.32\\
\cline{1-8}\lower0.35cm \hbox{{\vrule width 0pt height 1.0cm }}
$f_{2}^{D_s^+\to K_1^0(1270)}(q^2)$ &4.78&-1.42&1.37&$f_{2}^{D_s^+\to K_1^0(1400)}(q^2)$&-7.57 &-0.40&1.37\\
\cline{1-8}
\end{tabular}
\vspace{0.01cm} \caption{Parameters appearing in the fit function
for the form factors of the $D_{q}\to K_1(1270)\ell\nu$ and
$D_{q}\to K_1(1400)\ell\nu$ decays at $M_1^2=8~GeV^2$,
$M_2^2=6~GeV^2$ and $\theta_{K_1}=58^\circ$.}
\end{table}
\clearpage
\begin{table}[h]\label{tab:3}
\centering
\begin{tabular}{c|ccc||c|ccc}
\cline{2-4}\cline{6-8}&$a$&$b$&$m_{fit}$&&$a$&$b$&$m_{fit}$\\
\cline{1-8}\lower0.35cm \hbox{{\vrule width 0pt height 1.0cm }}
$f_{V}^{D\to K_1(1270)}(q^2)$ &5.48&-1.48 &1.23&$f_{V}^{D\to K_1(1400)}(q^2)$&2.81 &-0.54&1.23\\
\cline{1-8}\lower0.35cm \hbox{{\vrule width 0pt height 1.0cm }}
$f_{0}^{D\to K_1(1270)}(q^2)$ &-2.95&2.02 &1.33&$f_{0}^{D\to K_1(1400)}(q^2)$&-1.71 &1.22&1.33\\
\cline{1-8}\lower0.35cm \hbox{{\vrule width 0pt height 1.0cm }}
$f_{1}^{D\to K_1(1270)}(q^2)$ &0.61&-0.17 &1.25&$f_{1}^{D\to K_1(1400)}(q^2)$&0.35 &-0.12&1.25\\
\cline{1-8}\lower0.35cm \hbox{{\vrule width 0pt height 1.0cm }}
$f_{2}^{D\to K_1(1270)}(q^2)$ &3.90&-0.66 &1.29&$f_{2}^{D\to K_1(1400)}(q^2)$&2.10 &-0.28&1.29\\
\cline{1-8}\lower0.35cm \hbox{{\vrule width 0pt height 1.0cm }}
$f_{V}^{D_s^+\to K_1^0(1270)}(q^2)$ &7.23&-2.37 &1.27&$f_{V}^{D_s^+\to K_1^0(1400)}(q^2)$&2.10 &0.66&1.27\\
\cline{1-8}\lower0.35cm \hbox{{\vrule width 0pt height 1.0cm }}
$f_{0}^{D_s^+\to K_1^0(1270)}(q^2)$ &-4.27&2.83 &1.48&$f_{0}^{D_s^+\to K_1^0(1400)}(q^2)$&-2.43 &1.67&1.48\\
\cline{1-8}\lower0.35cm \hbox{{\vrule width 0pt height 1.0cm }}
$f_{1}^{D_s^+\to K_1^0(1270)}(q^2)$ &-0.80&0.14 &1.30&$f_{1}^{D_s^+\to K_1^0(1400)}(q^2)$&-0.54&0.15&1.30\\
\cline{1-8}\lower0.35cm \hbox{{\vrule width 0pt height 1.0cm }}
$f_{2}^{D_s^+\to K_1^0(1270)}(q^2)$ &7.05&0.28 &1.36&$f_{2}^{D_s^+\to K_1^0(1400)}(q^2)$&5.62 &-1.44&1.36\\
\cline{1-8}
\end{tabular}
\vspace{0.01cm} \caption{Parameters appearing in the fit function
for the form factors of the $D_{q}\to K_1(1270)\ell\nu$ and
$D_{q}\to K_1(1400)\ell\nu$ decays at $M_1^2=8~GeV^2$,
$M_2^2=6~GeV^2$ and $\theta_{K_1}=-37^\circ$.}
\end{table}
\begin{table}[h]\label{tab:3}
\centering
\begin{tabular}{c|ccc||c|ccc}
\cline{2-4}\cline{6-8}&$a$&$b$&$m_{fit}$&&$a$&$b$&$m_{fit}$\\
\cline{1-8}\lower0.35cm \hbox{{\vrule width 0pt height 1.0cm }}
$f_{V}^{D\to K_1(1270)}(q^2)$ &3.86&-0.91 &1.24&$f_{V}^{D\to K_1(1400)}(q^2)$&4.88 &-1.28&1.24\\
\cline{1-8}\lower0.35cm \hbox{{\vrule width 0pt height 1.0cm }}
$f_{0}^{D\to K_1(1270)}(q^2)$ &-2.17&1.49 &1.35&$f_{0}^{D\to K_1(1400)}(q^2)$&-2.57 &1.80&1.35\\
\cline{1-8}\lower0.35cm \hbox{{\vrule width 0pt height 1.0cm }}
$f_{1}^{D\to K_1(1270)}(q^2)$ &0.44&-0.10 &1.26&$f_{1}^{D\to K_1(1400)}(q^2)$&0.56 &-0.18&1.26\\
\cline{1-8}\lower0.35cm \hbox{{\vrule width 0pt height 1.0cm }}
$f_{2}^{D\to K_1(1270)}(q^2)$ &2.97&-0.61 &1.27&$f_{2}^{D\to K_1(1400)}(q^2)$&3.38 &-0.48&1.27\\
\cline{1-8}\lower0.35cm \hbox{{\vrule width 0pt height 1.0cm }}
$f_{V}^{D_s^+\to K_1^0(1270)}(q^2)$ &5.73&-2.15 &1.29&$f_{V}^{D_s^+\to K_1^0(1400)}(q^2)$&4.88 &-0.48&1.29\\
\cline{1-8}\lower0.35cm \hbox{{\vrule width 0pt height 1.0cm }}
$f_{0}^{D_s^+\to K_1^0(1270)}(q^2)$ &-3.14&2.07 &1.49&$f_{0}^{D_s^+\to K_1^0(1400)}(q^2)$&-3.70 &2.50&1.49\\
\cline{1-8}\lower0.35cm \hbox{{\vrule width 0pt height 1.0cm }}
$f_{1}^{D_s^+\to K_1^0(1270)}(q^2)$ &-0.58&0.08 &1.32&$f_{1}^{D_s^+\to K_1^0(1400)}(q^2)$&-0.78 &0.17&1.32\\
\cline{1-8}\lower0.35cm \hbox{{\vrule width 0pt height 1.0cm }}
$f_{2}^{D_s^+\to K_1^0(1270)}(q^2)$ &4.69&0.71 &1.35&$f_{2}^{D_s^+\to K_1^0(1400)}(q^2)$&7.84 &-1.20&1.35\\
\cline{1-8}
\end{tabular}
\vspace{0.01cm} \caption{Parameters appearing in the fit function
for the form factors of the $D_{q}\to K_1(1270)\ell\nu$ and
$D_{q}\to K_1(1400)\ell\nu$ decays at $M_1^2=8~GeV^2$,
$M_2^2=6~GeV^2$ and $\theta_{K_1}=-58^\circ$.}
\end{table}

At the end of this section, we would like to discuss the numeric values of the  differential decay rates as well as the   branching ratios for the considered semileptonic and nonleptonic transitions.

\section*{semileptonic}
 The dependence  of the longitudinal and transverse
components of the differential decay width for  the semileptonic
$D_{q}\to K_1 \ell \nu$ decays is shown in Figs. 9-20 at
$\theta_{K_1}=\pm37^\circ$. In these figures, the total decay widths
related to each decay are also depicted. To calculate  the branching
ratios of the semileptonic decays, we  Integrate Eq. (\ref{eq50})
over $q^2$ in the whole physical region and using the total mean
life-time $\tau_{D^0}=0.41~ps$, $\tau_{D^+}=1.04~ps$ and
$\tau_{D_s}=0.50~ps$ \cite{Yao}. The  values for the branching ratio
of these decays are obtained as presented in Table 7.
\begin{table}[h]\label{tab:3}
\centering
\begin{tabular}{c|cccc}
\cline{1-5}$\theta_{K_1}^{\circ}$&37&58&-37&-58\\
\cline{1-5}\lower0.25cm \hbox{{\vrule width 0pt height 1.0cm }}
$Br(D^0\to K_1^-(1270)\ell\nu)$&$[3.59\pm0.29$
&$1.03\pm0.10$&$5.34\pm0.21$
&$2.84\pm0.25]\times10^{-3}$\\
\cline{1-5}\lower0.25cm \hbox{{\vrule width 0pt height 1.0cm }}
$Br(D^+\to K_1^0(1270)\ell\nu)$ &$[9.47\pm0.45$
&$2.70\pm0.25$&$14.07\pm1.22$
&$7.57\pm0.35]\times10^{-3}$\\
\cline{1-5}\lower0.25cm \hbox{{\vrule width 0pt height 1.0cm }}
$Br(D_s^+\to K_1^0(1270)\ell\nu)$&$[7.84\pm0.41$
&$2.09\pm0.24$&$12.51\pm1.16$
&$6.91\pm0.32]\times10^{-4}$\\
\cline{1-5}\lower0.25cm \hbox{{\vrule width 0pt height 1.0cm }}
$Br(D^0\to K_1^-(1400)\ell\nu)$&$[1.09\pm0.10$
&$1.78\pm0.15$&$0.85\pm0.02$
&$1.20\pm0.11]\times10^{-3}$\\
\cline{1-5}\lower0.25cm \hbox{{\vrule width 0pt height 1.0cm }}
$Br(D^+\to K_1^0(1400)\ell\nu)$&$[2.93\pm0.25$
&$4.75\pm0.29$&$1.27\pm0.10$
&$3.20\pm0.27]\times10^{-3}$\\
\cline{1-5}\lower0.25cm \hbox{{\vrule width 0pt height 1.0cm }}
$Br(D_s^+\to K_1^0(1400)\ell\nu)$ &$[3.44\pm0.29$
&$5.88\pm0.34$&$1.49\pm0.13$
&$3.96\pm0.29]\times10^{-4}$\\
\cline{1-5}
\end{tabular}
\vspace{0.10cm} \caption{The values for the branching ratio of the semileptonic
$D_{q}\to K_1(1270) \ell \nu$ and $D_{q}\to K_1(1400) \ell
\nu$ decays  at different
values of the $\theta_{K_1}$.}
\end{table}
The errors in this Table are estimated by the variation of the Borel parameters
$M_1^2$
and $M_2^2$, the variation of the continuum thresholds $s_0$ and $s_0^\prime$ and uncertainties in the values of the
other input parameters.

\section*{nonleptonic}
For estimating the branching ratio of the  nonleptonic $D\rightarrow
K_1\pi$  decay, first the values of the form factors at
$q^2=m_{\pi}^{2}$ are calculated as  shown in Table 8.
\begin{table}[h]\label{tab:3}
\centering
\begin{tabular}{c|cccc||c|cccc}
\cline{2-5}\cline{7-10} $\theta_{K_1}^\circ$ &37&58&-37&-58&$\theta_{K_1}^\circ$ &37&58&-37&-58\\
\cline{1-10}\lower0.35cm \hbox{{\vrule width 0pt height 1.0cm }}
$f_{V}^{D\to K_1(1270)}$ &3.24&1.82 &4.04 &2.95 &$f_{V}^{D\to K_1(1400)}$&-3.45 &-4.42&2.30&3.65 \\
\cline{1-10}\lower0.35cm \hbox{{\vrule width 0pt height 1.0cm }}
$f_{0}^{D\to K_1(1270)}$ &-0.73&-0.42 &-0.91 &-0.67 &$f_{0}^{D\to K_1(1400)}$&0.70 &0.92&-0.47&-0.75 \\
\cline{1-10}\lower0.35cm \hbox{{\vrule width 0pt height 1.0cm }}
$f_{1}^{D\to K_1(1270)}$ &0.34&0.20 &0.45 &0.32 &$f_{1}^{D\to K_1(1400)}$&-0.36 &-0.49&0.25&0.41 \\
\cline{1-10}\lower0.35cm \hbox{{\vrule width 0pt height 1.0cm }}
$f_{2}^{D\to K_1(1270)}$ &2.67&1.55 &3.32 &2.49 &$f_{2}^{D\to K_1(1400)}$&-2.81&-3.65&1.87&3.03 \\
\cline{1-10}
\end{tabular}
\vspace{0.10cm} \caption{The values of the form factors of the
$D\to K_1(1270)$ and $D\to K_1(1400)$  for $M_1^2=8~GeV^2$,
$M_2^2=6~GeV^2$ at $q^2=m_\pi^2$ and different values of the mixing angle
$\theta_{K_1}$.}
\end{table}
Inserting these values in Eq. (\ref{eq35}) and using
$V_{ud}=0.97377\pm0.00027$ \cite{Yao}, $m_\pi=0.139~GeV$ and
$f_{\pi}=0.133~GeV$, we obtain the values for the branching ratio of
these decays as presented in Table 9. In comparison, we also include
the experimental values and upper limits in this Table. This Table
shows that for the $D^0\to K_1^-(1270)\pi^+$, $D^0\to
K_1^-(1400)\pi^+$ and $D^+\to K_1^0(1400)\pi^+$ cases, the different
values of mixing angle $\theta_{K_1}$ give the values of branching
ratios in good agreement with the experimental results but for
$D^+\to K_1^0(1270)\pi^+$ decay, the values of  the branching ratios
at different values of $\theta_{K_1}$ are about one order of
magnitude more than that of the experimental expectation.

\begin{table}[h]\label{tab:3}
\centering
\begin{tabular}{c|cccc|c}
\cline{1-6}$\theta_{K_1}^{\circ}$&37&58&-37&-58&Exp \cite{Yao}\\
\cline{1-6}\lower0.35cm \hbox{{\vrule width 0pt height 1.0cm }}
$Br(D^0\to K_1^{-}(1270)\pi^+)\times10^{-2}$ &$1.45\pm 0.11$&$0.75\pm 0.06 $&$2.26\pm 0.18$&$1.23\pm0.11$&$1.15\pm0.32$\\
\cline{1-6}\lower0.35cm \hbox{{\vrule width 0pt height 1.0cm }}
$Br(D^+\to K_1^0(1270)\pi^+)\times10^{-2}$ &$3.75\pm 0.29$&$1.23\pm 0.10 $&$5.85\pm 0.37$&$3.18\pm0.25$&$<0.7$\\
\cline{1-6}\lower0.35cm \hbox{{\vrule width 0pt height 1.0cm }}
$Br(D^0\to K_1^-(1400)\pi^+)\times10^{-2}$ &$0.60\pm 0.04$&$1.00\pm 0.12 $&$0.26\pm 0.02$&$0.73\pm0.04$&$<1.2$\\
\cline{1-6}\lower0.35cm \hbox{{\vrule width 0pt height 1.0cm }}
$Br(D^+\to K_1^0(1400)\pi^+)\times10^{-2}$ &$2.57\pm 0.21$&$3.63\pm 0.31 $&$1.71\pm 0.13$&$2.78\pm0.24$&$3.8\pm1.3$\\
\cline{1-6}
\end{tabular}
\vspace{0.10cm} \caption{The branching ratios of the nonleptonic
$D\to K_1(1270) \pi$ and $D\to K_1(1400)
 \pi$ decays  at
different values of $\theta_{K_1}$.}
\end{table}

In summary, we analyzed the semileptonic $D_{q}\to K_1 \ell\nu$
transition with $q=u, d, s$
 in the framework of the three--point QCD sum rules and
the nonleptonic $D\to K_1 \pi$ decay  within the factorization
approach. We calculated $D_{q}$ to $K_1(1270)$ and $K_1(1400)$ transition
form factors by separating the mixture of  the $K_1(1270)$ and
$K_1(1400)$ states. Using the transition form factors of the $D\to
K_1 $, we analyzed the nonleptonic $D\to K_1 \pi$ decay. We also
evaluated the decay amplitude and decay width of these decays in terms
of the transition form factors. The branching ratios of these decays
were also calculated at different values of the mixing angle
$\theta_{K_1}$. For the non leptonic case, a comparison of the results for the branching ratios with the existing experimental results was also made.
\section*{Acknowledgments}
Partial support of Shiraz university research council is
appreciated. K. A.  would like to thank T. M. Aliev and A. Ozpineci for
  their useful discussions and also TUBITAK, Turkish Scientific and Research
Council, for their partial financial support.

\clearpage
\begin{figure}\label{fig1}
\vspace*{-1cm}
\begin{center}
\includegraphics[width=8cm]{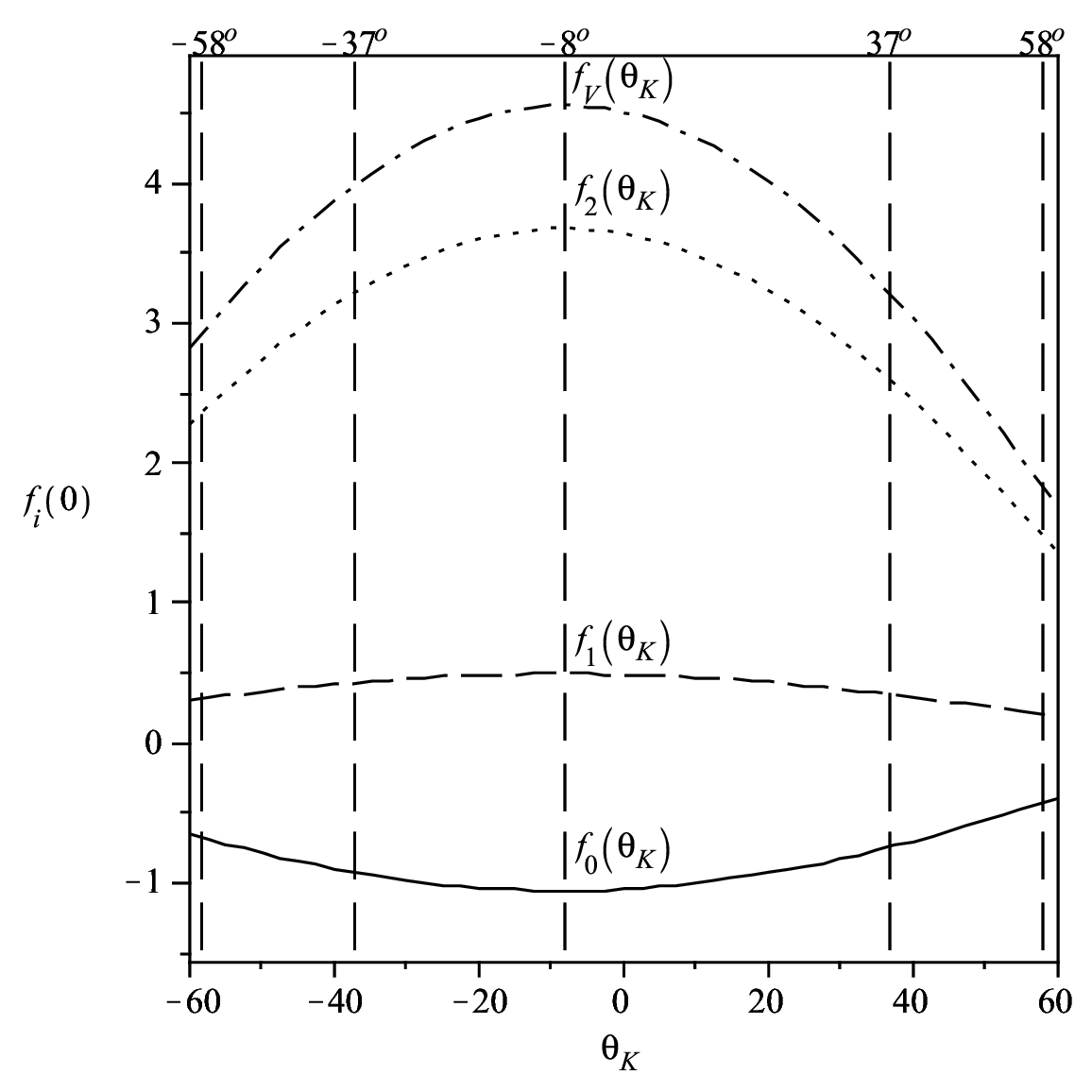}
\end{center}
\caption{The dependence of the form factors on $\theta_{K_1}$ at
$q^2=0$ for $D\to K_1(1270) \ell \nu$ decay.}
\end{figure}

\begin{figure}\label{fig1}
\vspace*{-1cm}
\begin{center}
\includegraphics[width=8cm]{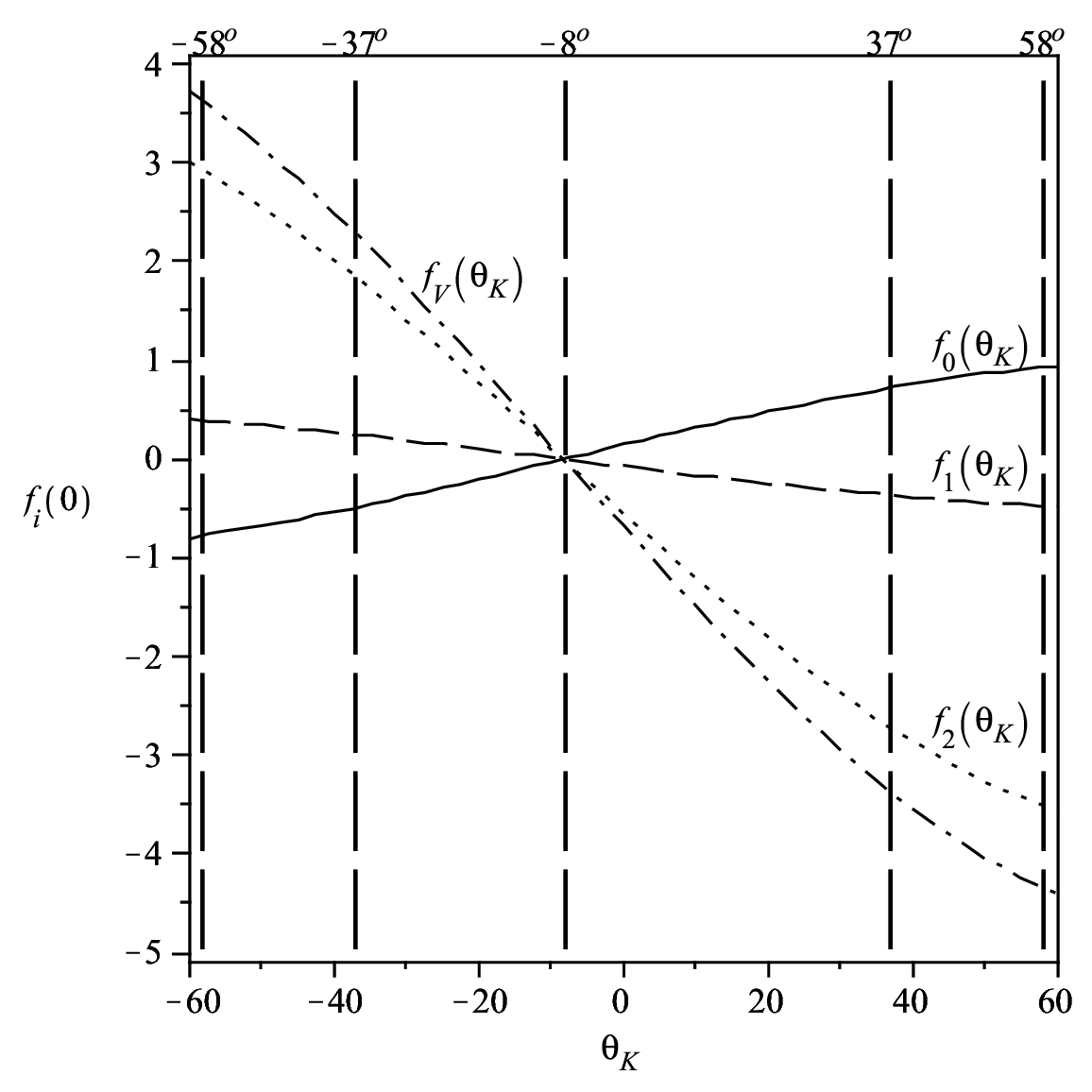}
\end{center}
\caption{The dependence of the form factors on $\theta_{K_1}$ at
$q^2=0$ for $D\to K_1(1400) \ell \nu$ decay.}
\end{figure}

\begin{figure}\label{fig1}
\vspace*{-1cm}
\begin{center}
\includegraphics[width=8cm]{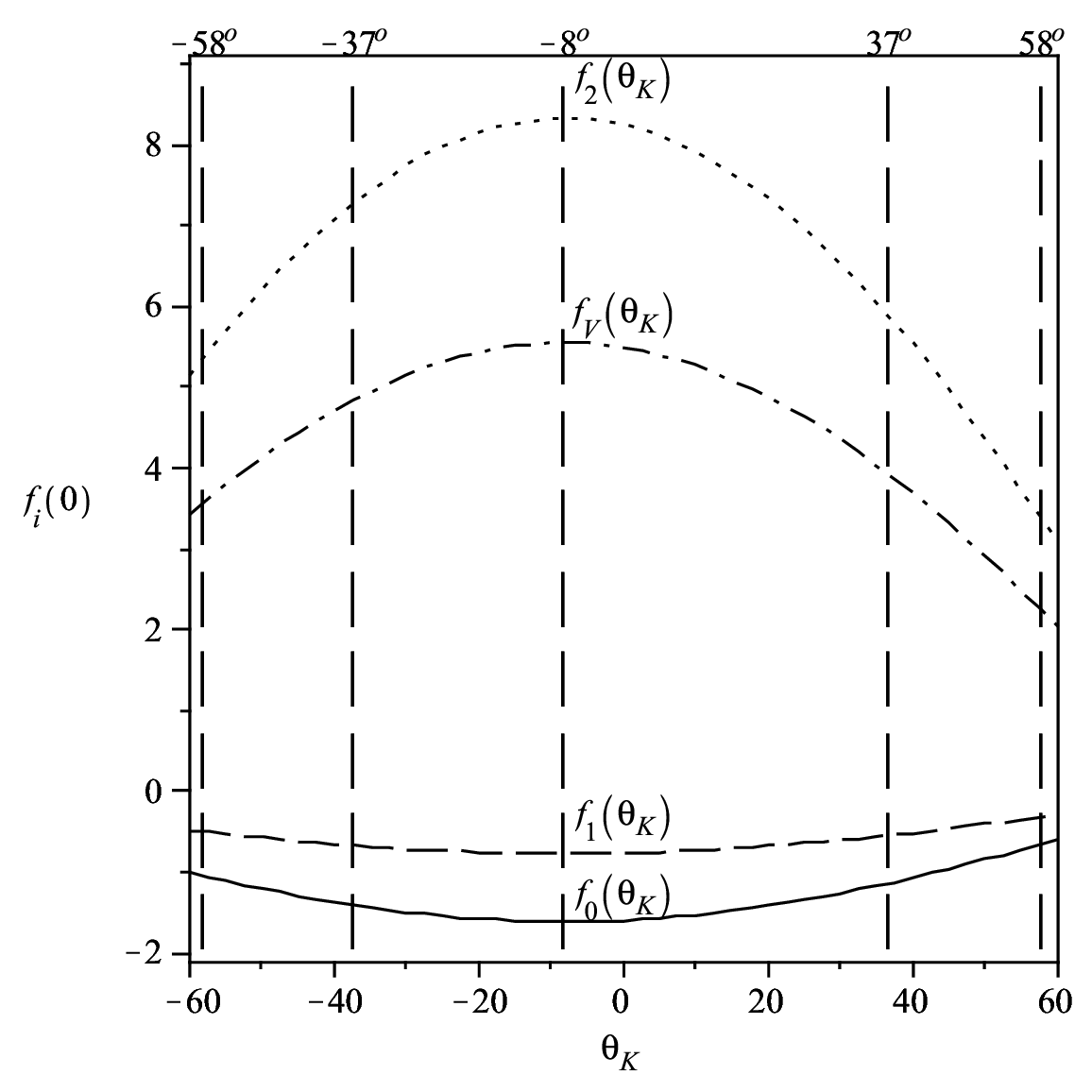}
\end{center}
\caption{The dependence of the form factors on $\theta_{K_1}$ at
$q^2=0$ for $D_s\to K_1(1270) \ell \nu$ decay.}
\end{figure}

\begin{figure}\label{fig1}
\vspace*{-1cm}
\begin{center}
\includegraphics[width=8cm]{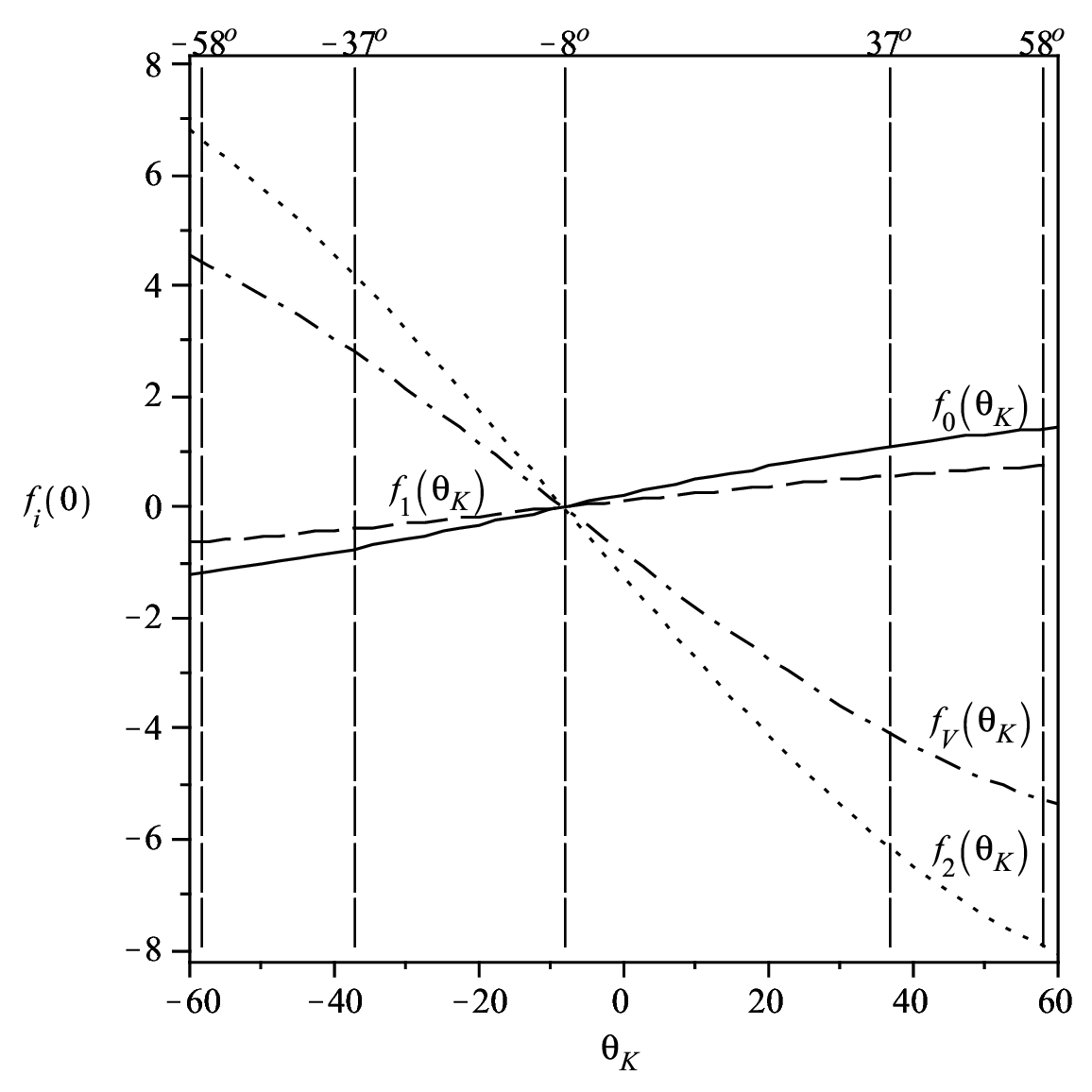}
\end{center}
\caption{The dependence of the form factors on $\theta_{K_1}$ at
$q^2=0$ for $D_s\to K_1(1400) \ell \nu$ decay.}
\end{figure}

\clearpage
\begin{figure}\label{fig1}
\vspace*{-1cm}
\begin{center}
\includegraphics[width=8cm]{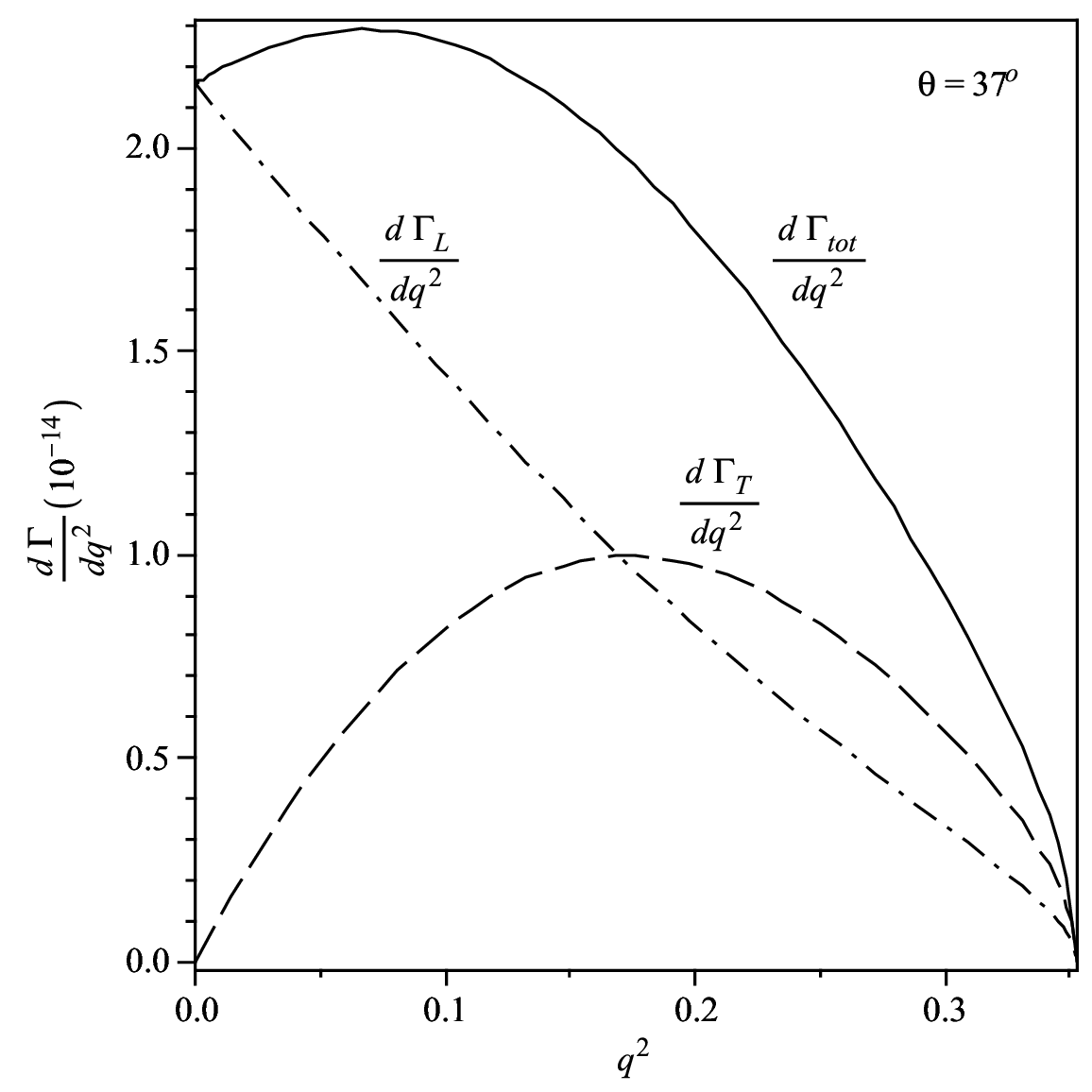}
\end{center}
\caption{The dependence of the $d\Gamma_{tot}/dq^2$,
$d\Gamma_{T}/dq^2$ and  $d\Gamma_{L}/dq^2$ on $q^2$ at
$\theta_{K_1}=37^\circ$ for $D^0\to K_1^-(1270) \ell\nu$.}
\end{figure}

\begin{figure}\label{fig1}
\vspace*{-1cm}
\begin{center}
\includegraphics[width=8cm]{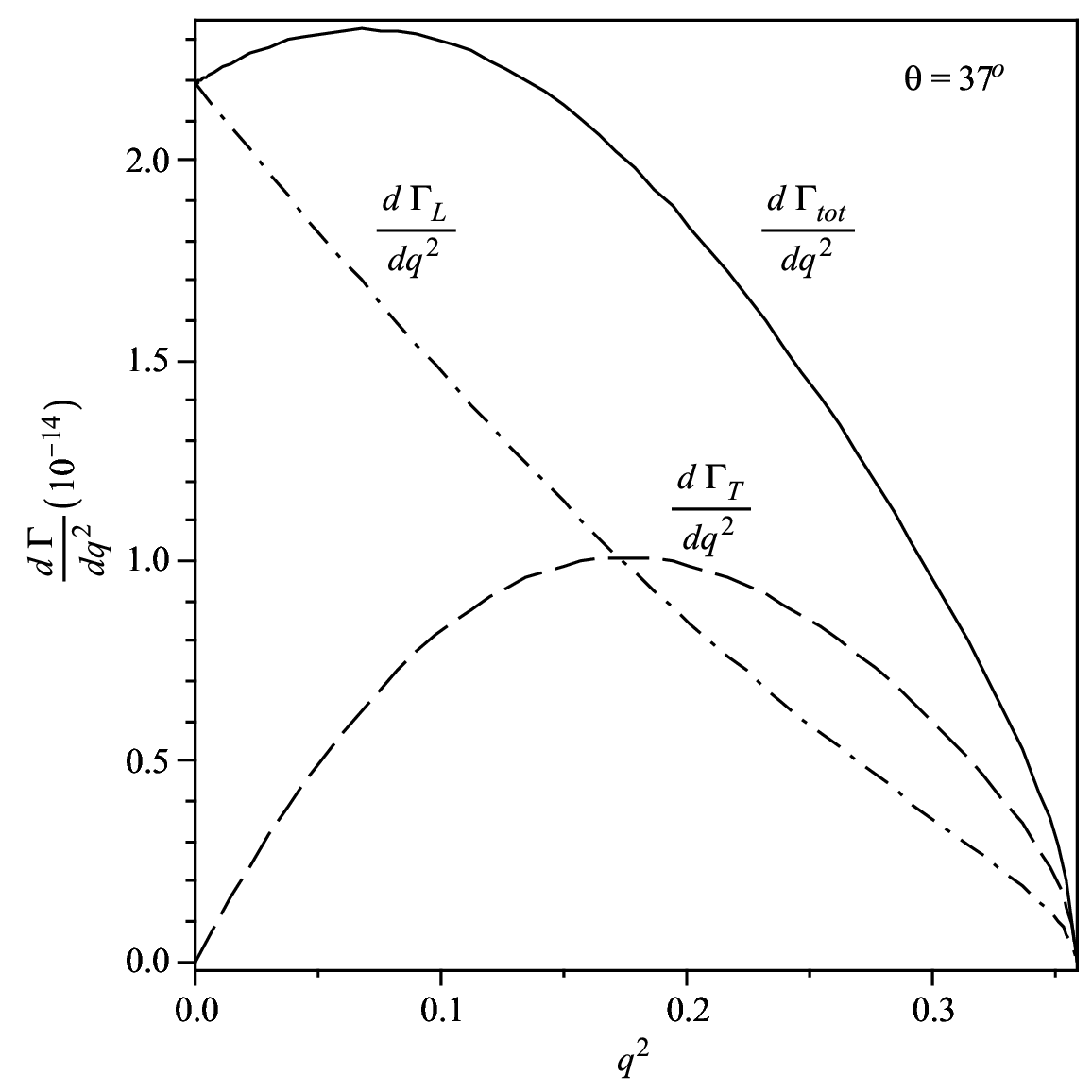}
\end{center}
\caption{The dependence of the $d\Gamma_{tot}/dq^2$,
$d\Gamma_{T}/dq^2$ and $d\Gamma_{L}/dq^2$ on $q^2$ at
$\theta_{K_1}=37^\circ$ for $D^+\to K_1^0(1270) \ell\nu$.}
\end{figure}

\begin{figure}\label{fig1}
\vspace*{-1cm}
\begin{center}
\includegraphics[width=8cm]{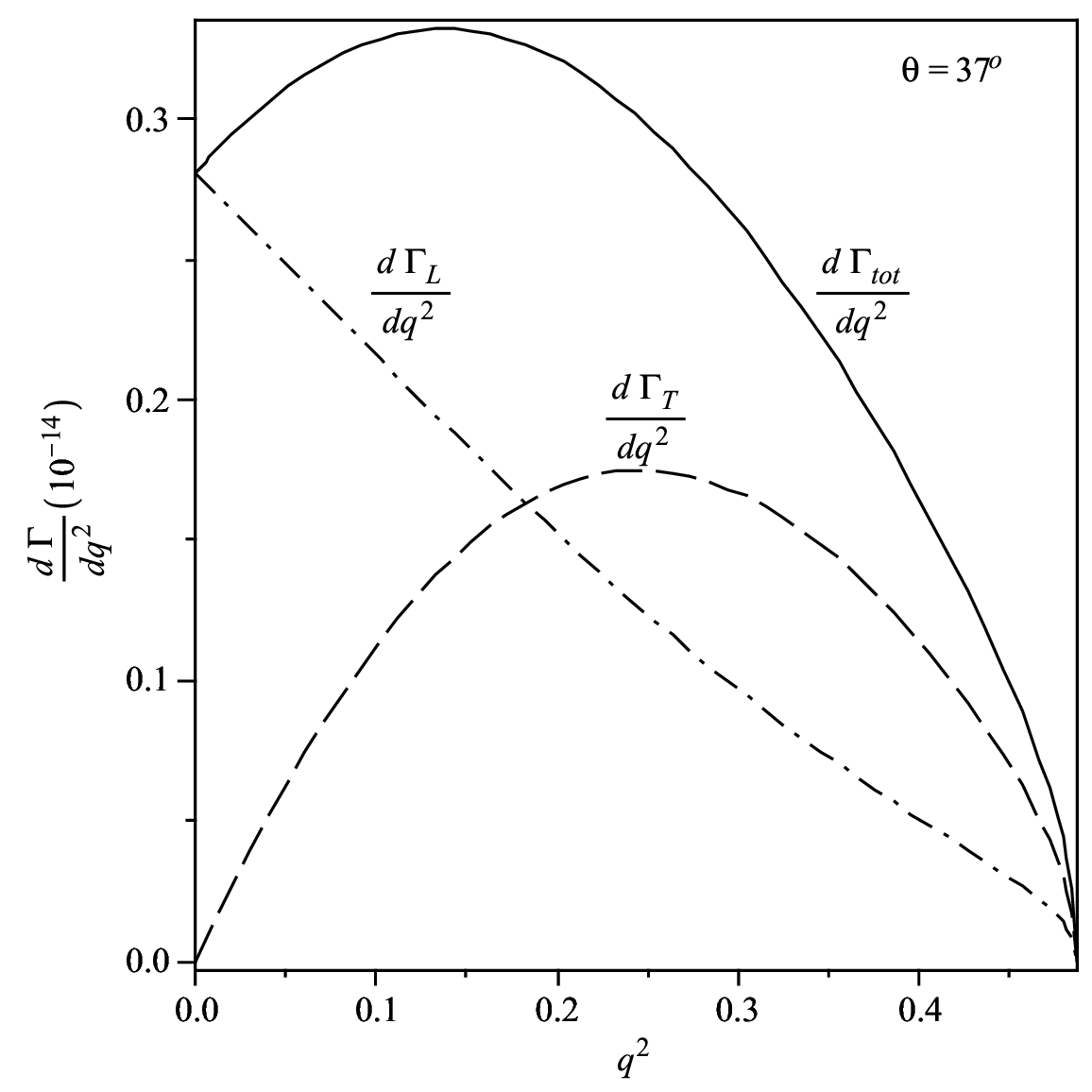}
\end{center}
\caption{The dependence of the $d\Gamma_{tot}/dq^2$,
$d\Gamma_{T}/dq^2$ and $d\Gamma_{L}/dq^2$ on $q^2$ at
$\theta_{K_1}=37^\circ$ for $D^+_s\to K_1^0(1270) \ell\nu$.}
\end{figure}

\begin{figure}\label{fig1}
\vspace*{-1cm}
\begin{center}
\includegraphics[width=8cm]{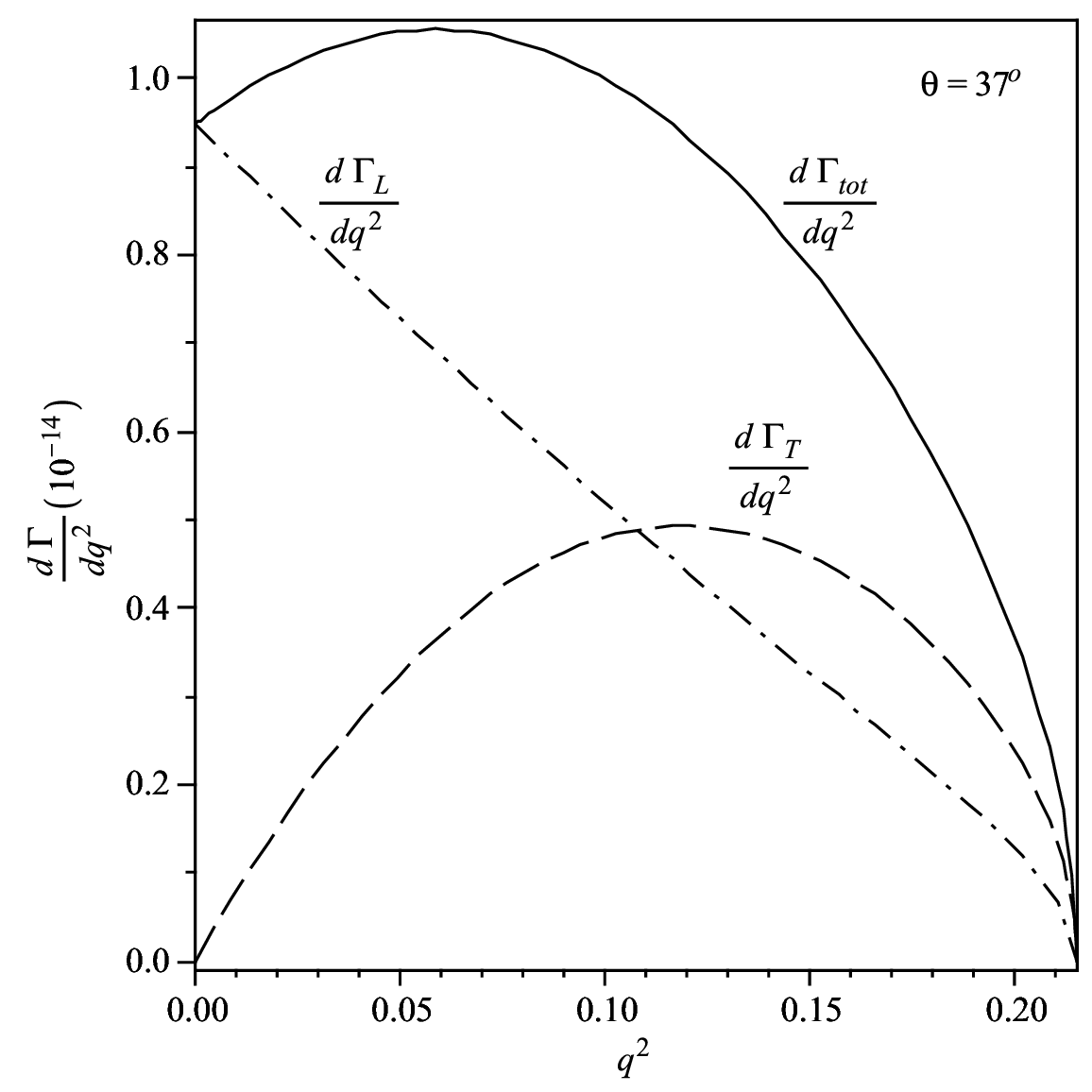}
\end{center}
\caption{The dependence of the $d\Gamma_{tot}/dq^2$,
$d\Gamma_{T}/dq^2$ and $d\Gamma_{L}/dq^2$ on $q^2$ at
$\theta_{K_1}=37^\circ$ for $D^0\to K_1^-(1400) \ell\nu$.}
\end{figure}

\begin{figure}\label{fig1}
\vspace*{-1cm}
\begin{center}
\includegraphics[width=8cm]{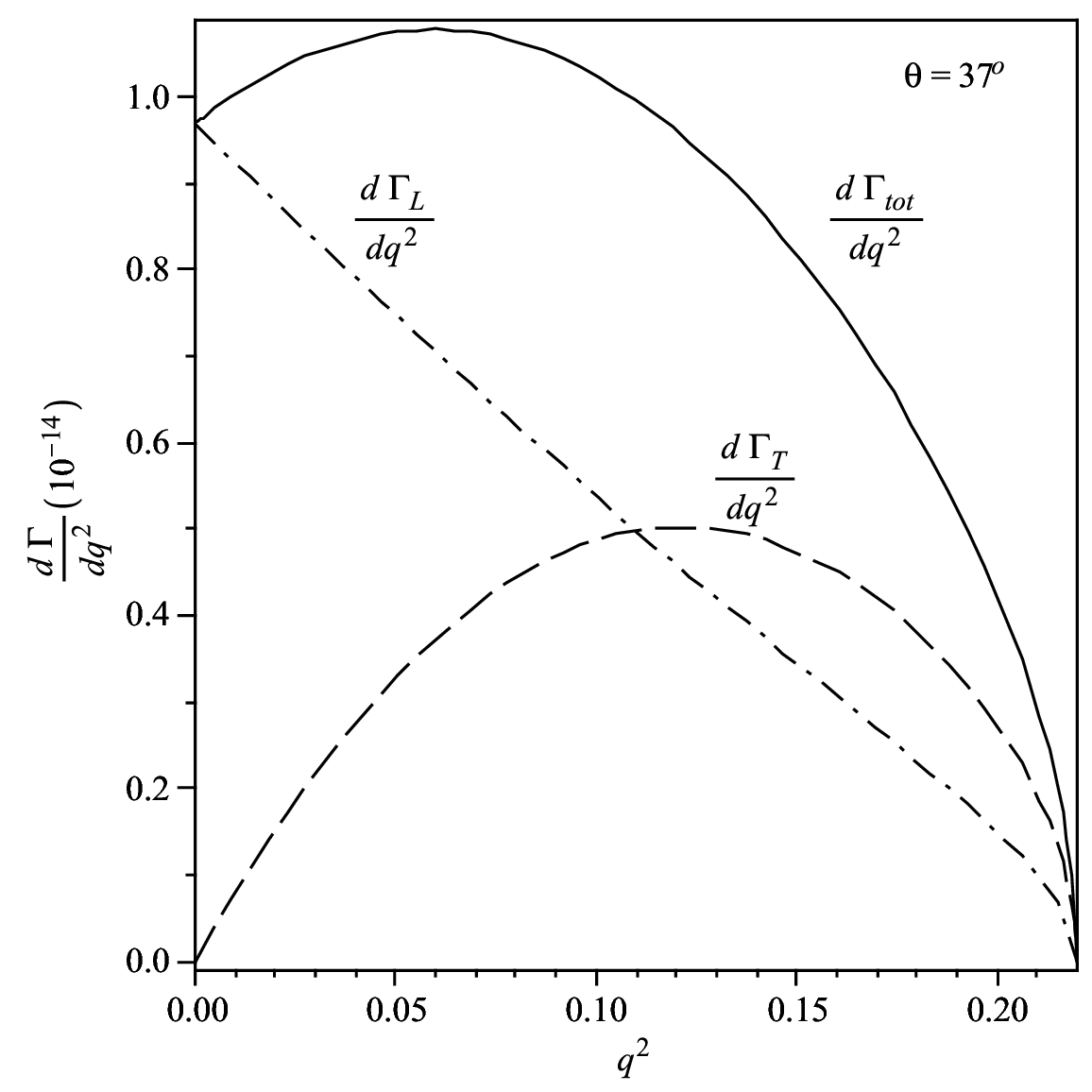}
\end{center}
\caption{The dependence of the $d\Gamma_{tot}/dq^2$,
$d\Gamma_{T}/dq^2$ and $d\Gamma_{L}/dq^2$ on $q^2$ at
$\theta_{K_1}=37^\circ$ for $D^+\to K_1^0(1400) \ell\nu$.}
\end{figure}

\begin{figure}\label{fig1}
\vspace*{-1cm}
\begin{center}
\includegraphics[width=8cm]{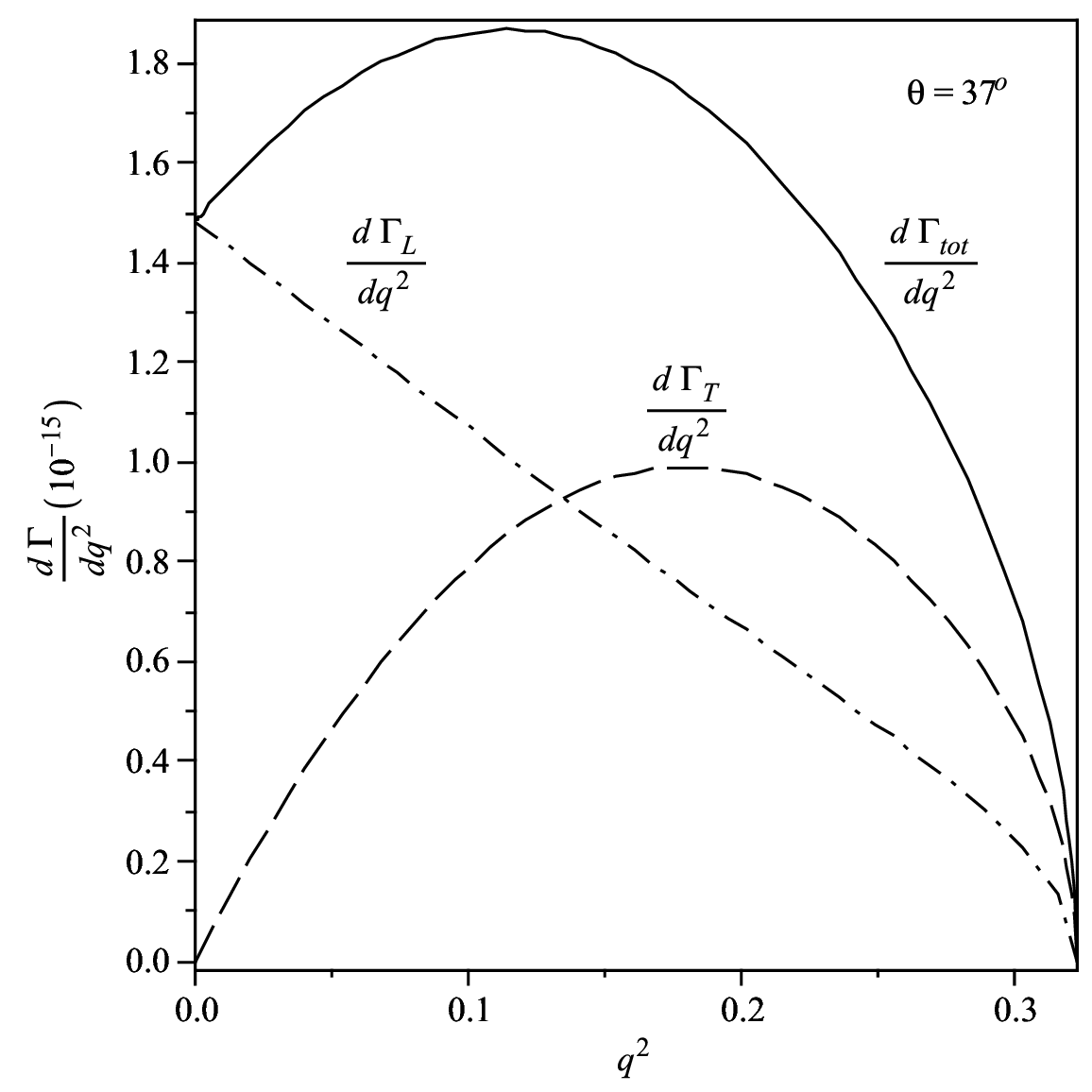}
\end{center}
\caption{The dependence of the $d\Gamma_{tot}/dq^2$,
$d\Gamma_{T}/dq^2$ and $d\Gamma_{L}/dq^2$ on $q^2$ at
$\theta_{K_1}=37^\circ$ for $D^+_s\to K_1^0(1400) \ell\nu$.}
\end{figure}

\clearpage
\begin{figure}\label{fig1}
\vspace*{-1cm}
\begin{center}
\includegraphics[width=8cm]{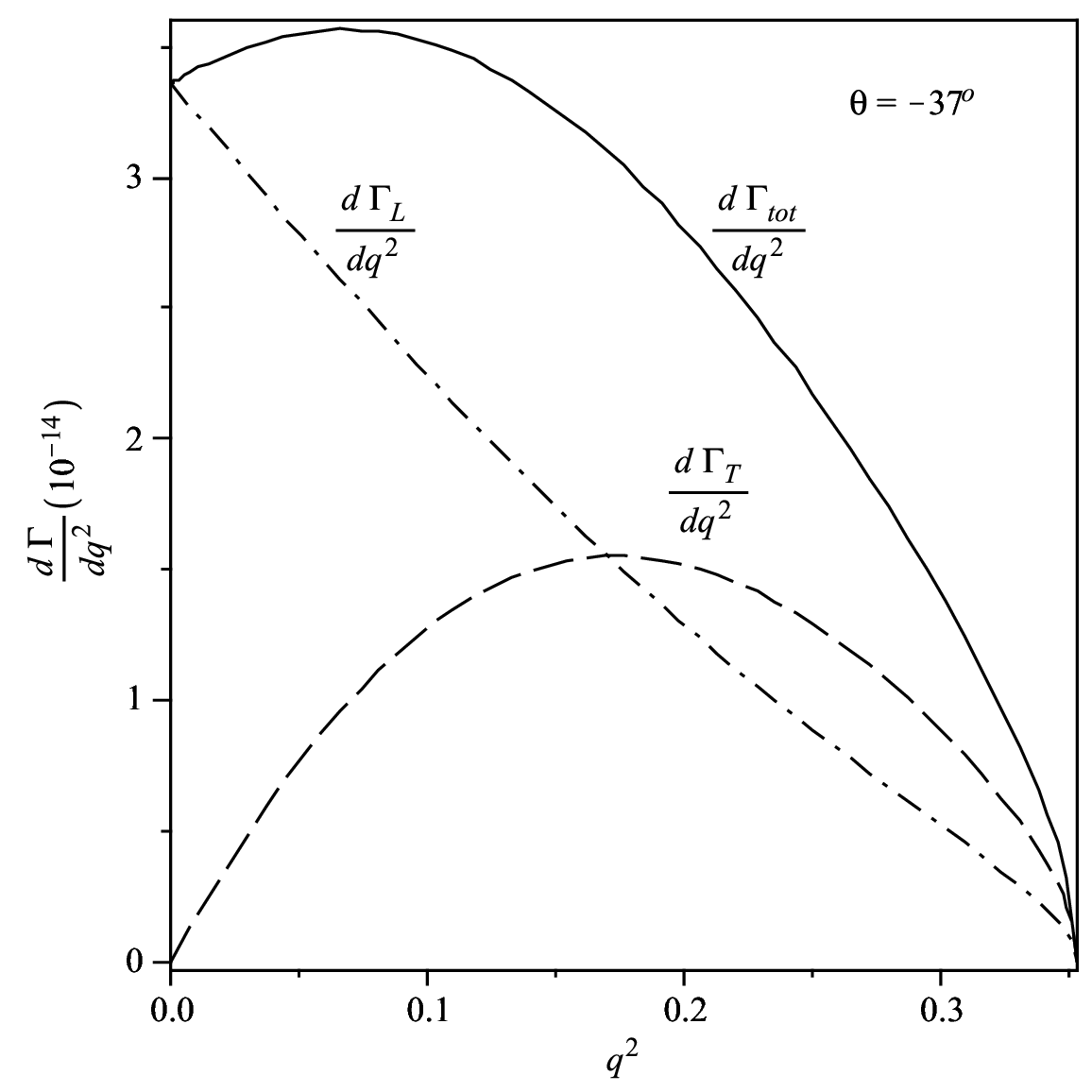}
\end{center}
\caption{The dependence of the $d\Gamma_{tot}/dq^2$,
$d\Gamma_{T}/dq^2$ and $d\Gamma_{L}/dq^2$ on $q^2$ at
$\theta_{K_1}=-37^\circ$ for $D^0\to K_1^-(1270) \ell\nu$.}
\end{figure}

\begin{figure}\label{fig1}
\vspace*{-1cm}
\begin{center}
\includegraphics[width=8cm]{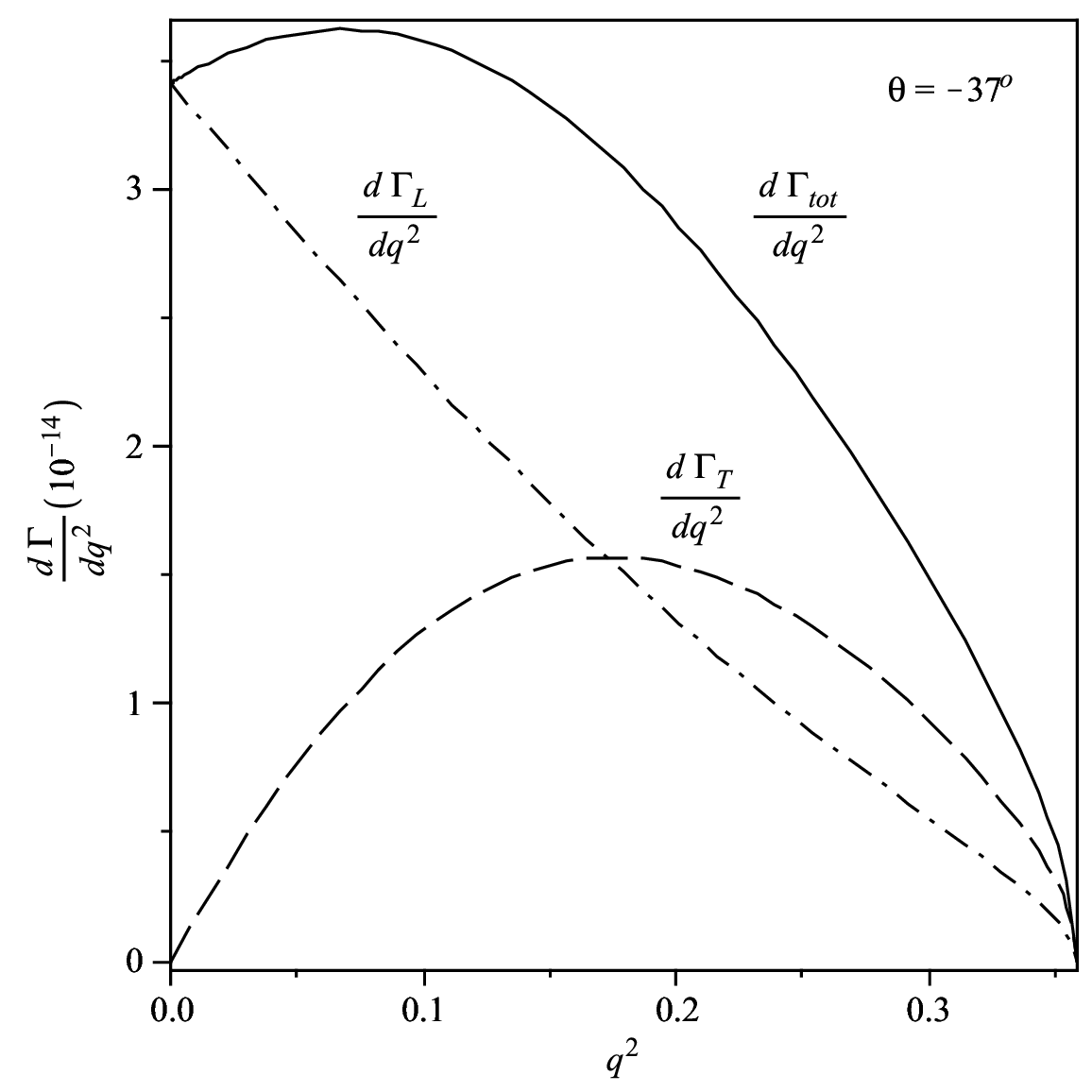}
\end{center}
\caption{The dependence of the $d\Gamma_{tot}/dq^2$,
$d\Gamma_{T}/dq^2$ and $d\Gamma_{L}/dq^2$ on $q^2$ at
$\theta_{K_1}=-37^\circ$ for $D^+\to K_1^0(1270) \ell\nu$.}
\end{figure}

\begin{figure}\label{fig1}
\vspace*{-1cm}
\begin{center}
\includegraphics[width=8cm]{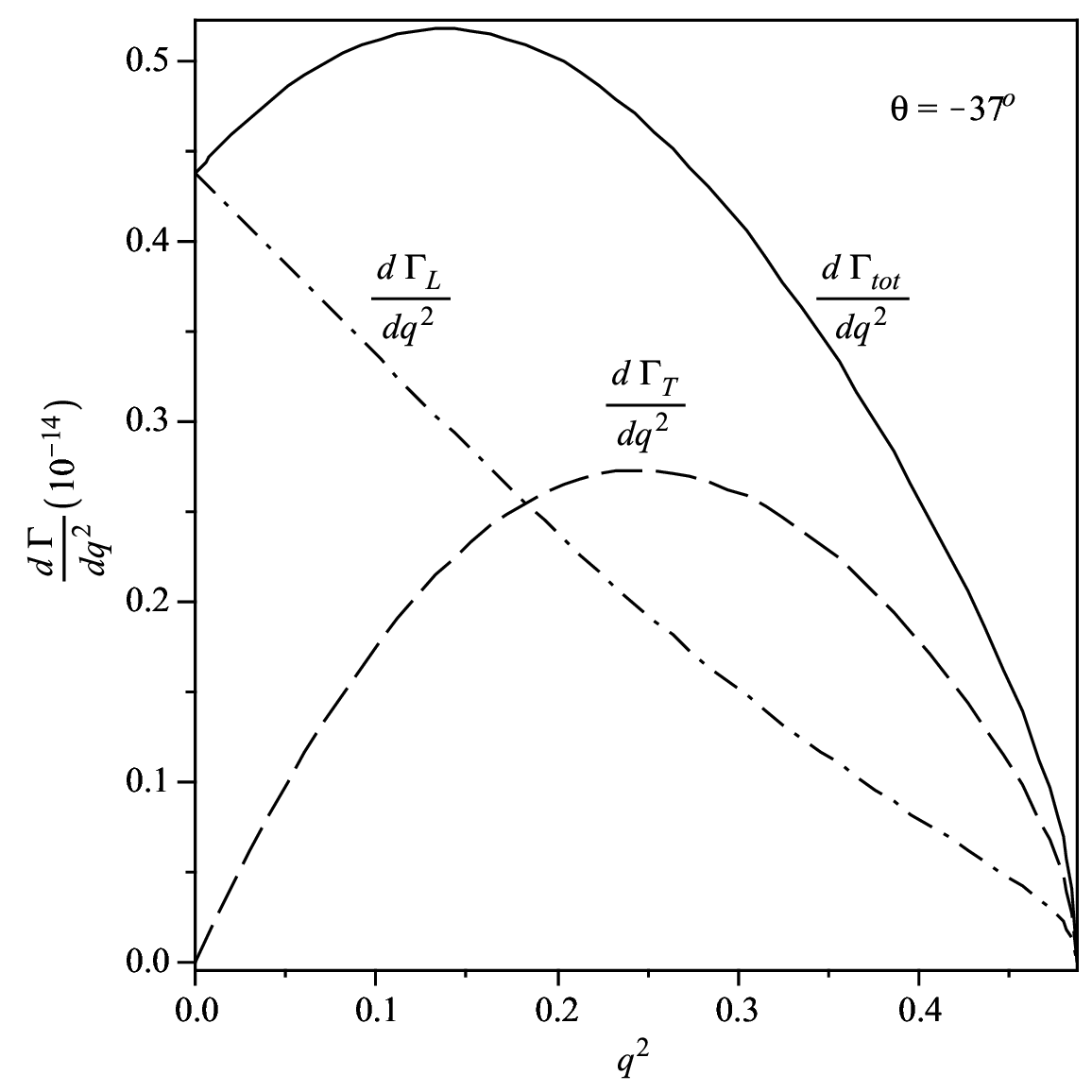}
\end{center}
\caption{The dependence of the $d\Gamma_{tot}/dq^2$,
$d\Gamma_{T}/dq^2$ and $d\Gamma_{L}/dq^2$ on $q^2$ at
$\theta_{K_1}=-37^\circ$ for $D^+_s\to K_1^0(1270) \ell\nu$.}
\end{figure}

\begin{figure}\label{fig1}
\vspace*{-1cm}
\begin{center}
\includegraphics[width=8cm]{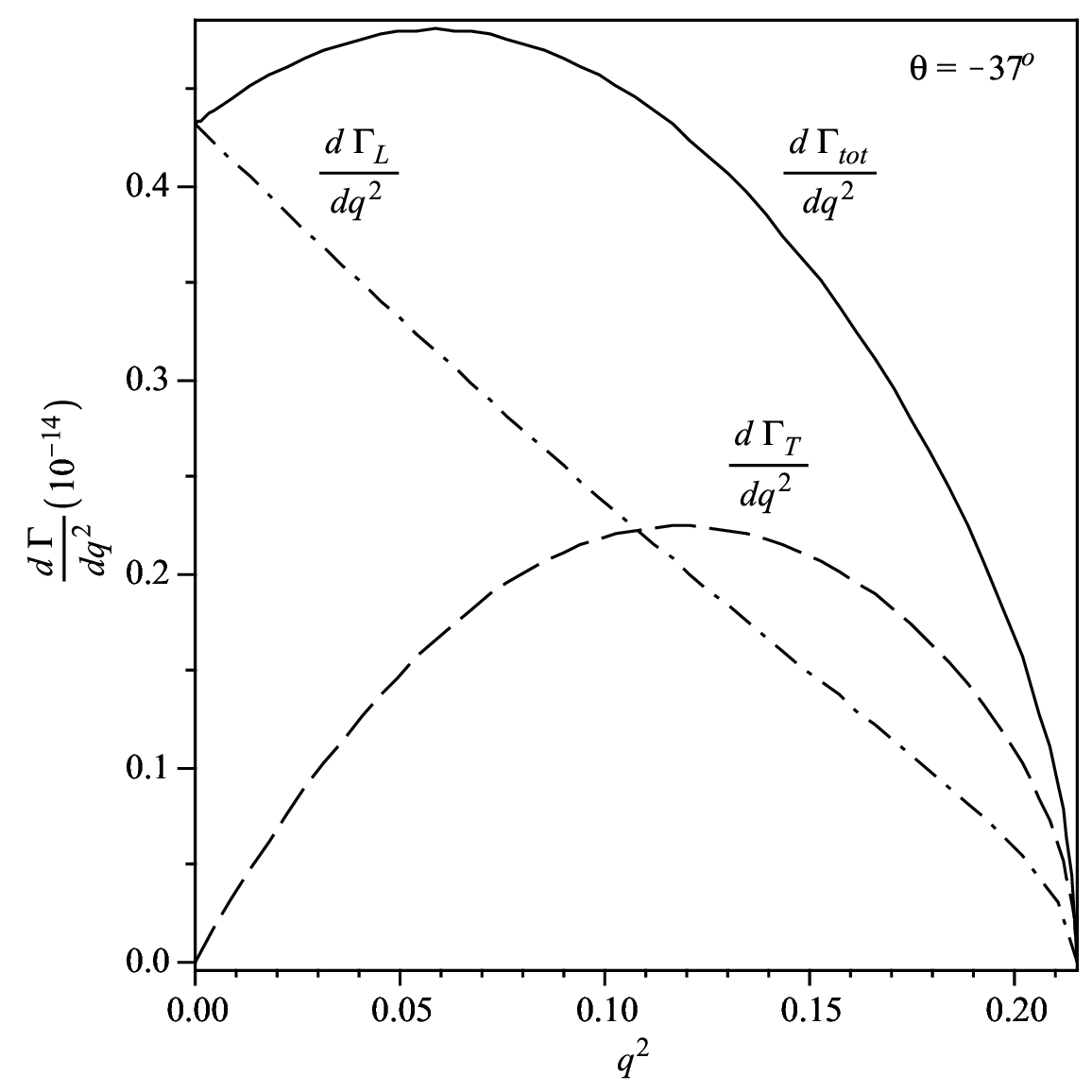}
\end{center}
\caption{The dependence of the $d\Gamma_{tot}/dq^2$,
$d\Gamma_{T}/dq^2$ and $d\Gamma_{L}/dq^2$ on $q^2$ at
$\theta_{K_1}=-37^\circ$ for $D^0\to K_1^-(1400) \ell\nu$.}
\end{figure}

\begin{figure}\label{fig1}
\vspace*{-1cm}
\begin{center}
\includegraphics[width=8cm]{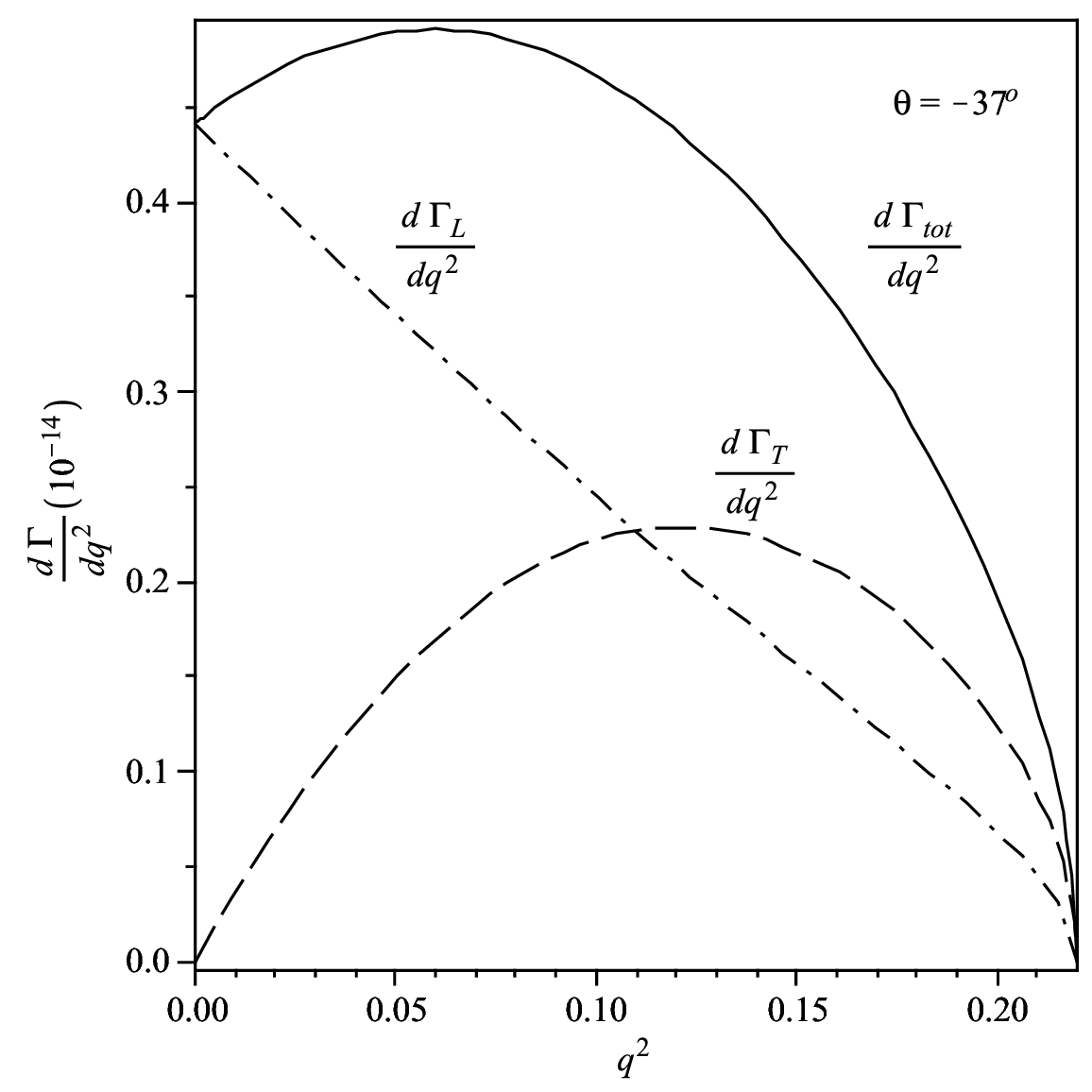}
\end{center}
\caption{The dependence of the $d\Gamma_{tot}/dq^2$,
$d\Gamma_{T}/dq^2$ and $d\Gamma_{L}/dq^2$ on $q^2$ at
$\theta_{K_1}=-37^\circ$ for $D^+\to K_1^0(1400) \ell\nu$.}
\end{figure}

\begin{figure}\label{fig1}
\vspace*{-1cm}
\begin{center}
\includegraphics[width=8cm]{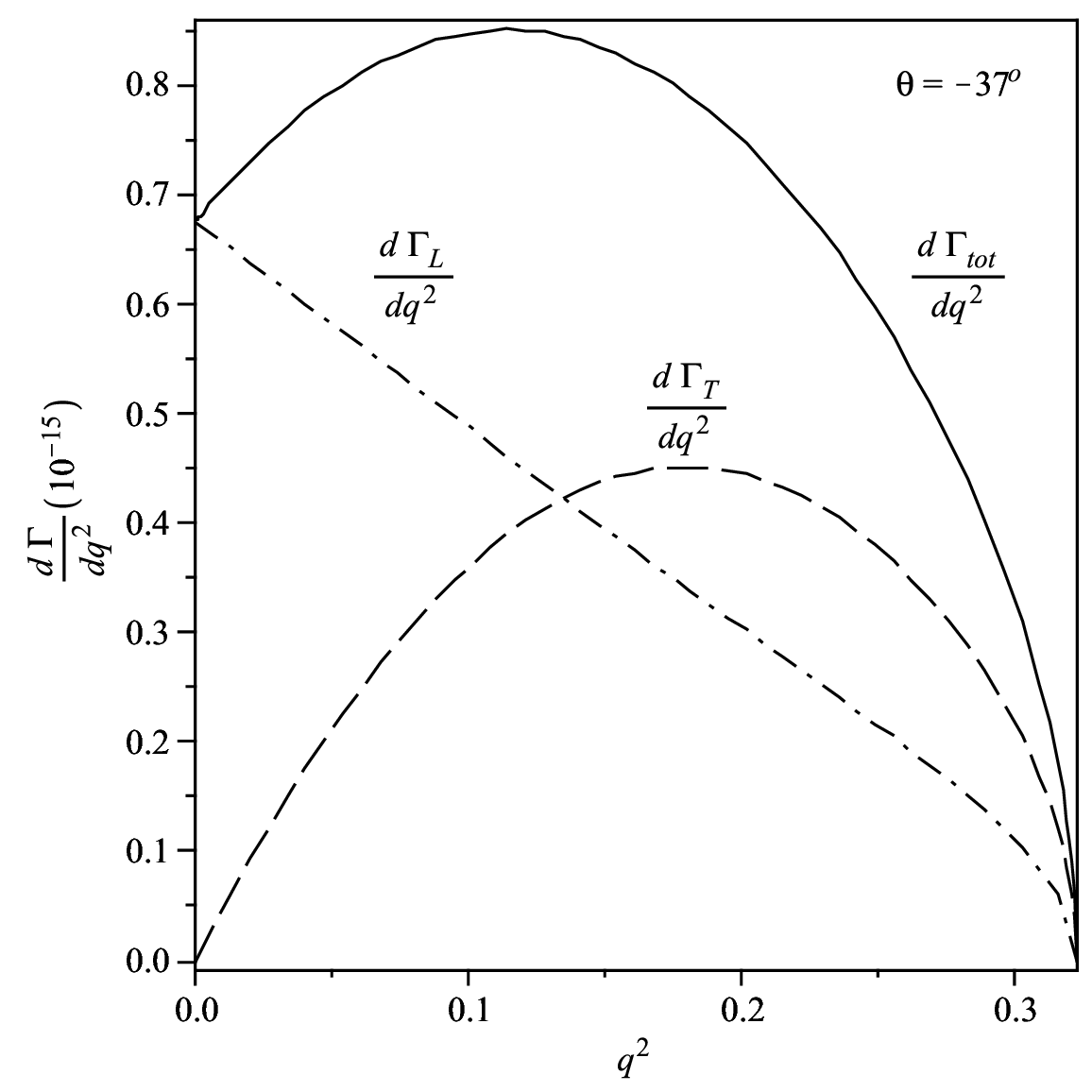}
\end{center}
\caption{The dependence of the $d\Gamma_{tot}/dq^2$,
$d\Gamma_{T}/dq^2$ and $d\Gamma_{L}/dq^2$ on $q^2$ at
$\theta_{K_1}=-37^\circ$ for $D^+_s\to K_1^0(1400) \ell\nu$.}
\end{figure}

\end{document}